\documentclass{aa}  
\usepackage{amssymb}
\usepackage{amsfonts}
\usepackage{amsmath}
\usepackage{textcomp}
\usepackage{latexsym}

\usepackage{graphicx}
\usepackage{natbib}
\usepackage{txfonts}
\def\apjl{ApJ~Lett.}          
\def\apjs{ApJS}               
\def\aap{A\&A}                

\begin{document}
   \title{The $s$-process in stellar population synthesis:\\
     a new approach to understanding AGB stars}
   \titlerunning{$S$-process and population synthesis of AGB stars}

   \author{A. Bona\v{c}i\'c Marinovi\'c,
    R.G. Izzard, M. Lugaro, O.R. Pols
          }
   \authorrunning{Bona\v{c}i\'c Marinovi\'c et al.}
   \offprints{A. Bona\v{c}i\'c Marinovi\'c}

   \institute{Sterrekundig Instituut Utrecht (SIU), Universiteit Utrecht,
     P.O. Box 80000, NL-3508 TA Utrecht, The Netherlands.\\
              \email{bonacic@astro.uu.nl}
             }

   \date{Received December 4, 2006; accepted March 5, 2007}

 
  \abstract
   {Thermally pulsating asymptotic giant branch (AGB) stars are
     the main producers of slow neutron capture ($s$-) process
     elements, but there are still large uncertainties
     associated with the formation of the main neutron source,
     $^{13}$C, and with the physics of these stars in general.
     Observations of $s$-process element enhancements in stars
     can be used as constraints on theoretical models.
   }
   {For the first time we apply stellar population
     synthesis to the problem of $s$-process nucleosynthesis in AGB
     stars, in order to derive constraints on
     free parameters describing
     the physics behind the third dredge-up
     and the properties of the neutron source.
  }
   {We utilize a rapid evolution and nucleosynthesis code to 
     synthesize different populations of $s$-enhanced stars, and
     compare them to their observational counterparts to find
     out for which values of the free parameters in the code the 
     synthetic populations
     fit best to the observed populations. These free parameters
     are the amount of third dredge-up, the minimum core mass for
     third dredge-up, the effectiveness of $^{13}$C as a source of
     neutrons and the size in mass of the $^{13}$C pocket.
    }
   {We find that galactic disk objects are reproduced by a spread
     of a factor of two in the effectiveness of the $^{13}$C neutron
     source. Lower metallicity objects can be
     reproduced only by lowering by at least a factor of 3 the
     average value of the effectiveness of the $^{13}$C neutron source
     needed for the galactic disk objects.
     Using observations of $s$-process elements in post-AGB stars
     as constraints we find that dredge-up has to start at a lower
     core mass than predicted by current theoretical models, that it has
     to be substantial ($\lambda\gtrsim0.2$) in stars with mass
     $M\lesssim1.5$ M$_{\odot}$ and
     that the mass of the $^{13}$C pocket must be about
     1/40 that of the intershell region.
   }
   {}

   \maketitle

\section{Introduction}
About half of the elements heavier than iron are produced by slow
neutron captures ($s$-process) in the deep layers of thermally
pulsating (TP-) asymptotic giant branch (AGB) stars, which represent
the late evolutionary phase of objects with initial mass $M\lesssim$ 8
M$_{\odot}$.  These stars have degenerate carbon-oxygen cores
surrounded by two burning shells, and undergo a cyclic instability of
the helium-burning shell known as thermal pulses. $S$-process elements
are believed to be synthesized in the thin helium-rich layer
between the helium-burning and hydrogen-burning shells known as the
inter-shell region, or \emph{intershell}. After each thermal pulse the
hydrogen-rich convective envelope may penetrate into the intershell
and bring the synthesized heavy elements, together with the products
of helium burning (mainly carbon), to the stellar surface in a
phenomenon called \emph{third dredge-up} (TDU). Strong stellar winds
gradually ablate the envelope and enrich the interstellar medium with
the nucleosynthesis products. Many of the physical processes taking
place in TP-AGB stars are still highly uncertain, such as how the main
source of free neutrons required for the $s$-process is produced (see
below), under which conditions TDU occurs and how efficient it is, and
the strength and driving mechanism of mass loss. Many of these
uncertainties are related to our poor understanding of mixing
processes in stars. For a recent review on AGB stars see
\citet{2005ARA&A..43..435H} and for a discussion on uncertainties
in AGB stellar models see, e.g., \citet{2003ApJ...586.1305L}.

The synthesis of $s$-process elements requires a source of free
neutrons that can be captured by iron seeds to build up heavier nuclei
along the isotopic stability valley. During the period of quiescent
hydrogen shell burning in between thermal pulses, the inter-pulse
period, the conditions in the intershell (temperature, density and
helium abundance) favour the ${\rm ^{13}C}(\alpha,n)^{16}{\rm O}$
reaction. For this reaction to be an effective source of free neutrons
a $^{13}$C-rich region or \emph{pocket} in the intershell is
needed. The production of $^{13}$C takes place by means of the
${\rm ^{12}C}(p,\gamma){\rm ^{13}N}(\beta^+\nu){\rm ^{13}C}$
reaction.  It is assumed that the
protons needed for this reaction penetrate the top of the intershell
after TDU has occurred, as a result of a mixing process of as yet
uncertain nature: semi-convection \citep{1988ApJ...333L..25H},
hydrodynamical overshooting \citep{1997A&A...324L..81H}, rotation
\citep{1999A&A...346L..37L, 2003ApJ...593.1056H} and gravity waves
\citep{2003MNRAS.340..722D} have all been proposed.
As a consequence large uncertainties exist in the size in mass of
the $^{13}$C pocket and the $^{13}$C abundance profile inside it.
In addition, due to the existence of several neutron poisons,
especially $^{14}$N, not all free neutrons
released by the ${\rm ^{13}C}(\alpha,n)^{16}{\rm O}$ reaction
participate in the $s$-process, so the effectiveness of $^{13}$C as a
neutron source for the $s$-process is also very uncertain.
Another neutron burst is also released in the convective
intershell during the thermal pulses due to the
${\rm ^{22}Ne}(\alpha,n)^{25}{\rm Mg}$ reaction, but
the activation of this reaction is only marginal, which
makes its contribution to the neutron exposure smaller than that of
the ${\rm ^{13}C}(\alpha,n)^{16}{\rm O}$ reaction (see
\citealt{1999ARA&A..37..239B} for a review of $s$-process
nucleosynthesis in AGB stars).

Among stars that are enhanced in $s$-process elements ($s$-enhanced
stars) one distinguishes intrinsic and extrinsic objects.  Intrinsic
$s$-enhanced stars are typically late-type giants of spectral class S
and C that show over-abundances of $s$-process elements (see, e.g.,
\citealp{1990ApJS...72..387S,2002ApJ...579..817A}),
including the radioactive element Tc \citep{1952ApJ...116...21M}.
With a half-life
of a few million years, the presence of Tc lines indicates that the
synthesis of $s$-process elements happened recently and in
situ. Therefore intrinsic $s$-enhanced stars must be objects in the
TP-AGB phase or post-AGB phase, which have recently experienced
thermal pulses and TDU. The study of elemental abundances in these
stars gives clues on the uncertain physical phenomena that take place
in their interiors.  Extrinsic $s$-enhanced stars also show
over-abundances of $s$-process elements, but no Tc lines are present
\citep{1993A&A...271..463J}.  This indicates that
Tc has decayed since the $s$-process elements were synthesized. In
addition, extrinsic stars are often observed to be in an evolution
phase earlier than the TP-AGB, i.e., they are (sub)giants or main
sequence stars. This implies that the $s$-process elements were not
produced in situ, but that they have been accreted from a more massive
TP-AGB companion star. These stars can therefore act as probes for
studying mass transfer processes in binaries and for tracing the
nucleosynthesis that occurred in their companion stars.

\citet{2001ApJ...557..802B} carried out a detailed comparison
between observations
of $s$-process enhanced stars and model predictions for the
$s$-process in single AGB stars of different metallicities.  They
found that a large spread (a factor of $\sim20$) in the effectiveness
(see $\S$\ref{c13eff} for a definition of this parameter) of the main
neutron source $^{13}$C was needed to match spectroscopic observations
at a given metallicity. Their analysis is based on a small set of
initial stellar masses and they have taken into consideration only the
final abundances of their stars. In this paper we improve on their
analysis by comparing observations of different types of stars
belonging to the AGB family to stellar population synthesis models
computed with the inclusion of the $s$-process.  Our use of a rapid
synthetic evolution code allows us to study a large set of finely
spaced initial masses and metallicities and to trace the complete AGB
evolution of each star. Using this method we are able to put much
tighter constraints on the effectiveness of $^{13}$C as a neutron
source. In addition, by comparing our models to post-AGB observations
we also put constraints on the minimum core-mass for TDU, the TDU
efficiency and the size of the $^{13}$C pocket.

The AGB stellar models and the associated free parameters are
described in $\S$\ref{agbmodels}. Our stellar population synthesis
method is described in $\S$\ref{psyn} and the results are presented in
$\S$\ref{results}. In $\S$\ref{disc} we discuss our results and draw
our conclusions.

\section{The AGB models}
\label{agbmodels}
Our TP-AGB star models are calculated with a modified version of
the rapid evolutionary code by \citet{2000MNRAS.315..543H} and
\citet{2004MNRAS.350..407I}.
We limit ourselves to a brief overview of the ingredients
of this code and a discussion of the most important free parameters.
A detailed description of our modifications and improvements
can be found in Appendix \ref{apa}.
Stellar evolution and nucleosynthesis are modelled synthetically,
i.e., by means of
analytical fits to the detailed evolutionary models of
\citet{2002PASA...19..515K} and \citet{2004MNRAS.352..984S}, and
in some cases by tabular
interpolation of these detailed models.
Fits and interpolations depend on global
stellar parameters, such as metallicity, mass
and core-mass. This synthetic approach makes the calculation
of the evolution very fast.
The $s$-process element nucleosynthesis is calculated
by interpolating results based on detailed models by
\citet{1998ApJ...497..388G}, as described below.
With these models we are able to follow the chemical abundances
of stars as a function of time, from their initial
abundances at the beginning of the main sequence, for which we use
those of \citet{1989GeCoA..53..197A}, to their chemically
enhanced abundances of the latest evolutionary TP-AGB and
post-AGB phases.

Many uncertainties
exist in the detailed evolutionary models, which we treat as free
parameters. In the following subsections we discuss the
free parameters that are most relevant for stellar chemical
evolution.

\subsection{Minimum core-mass for third dredge-up}
Detailed models of TP-AGB stellar evolution which
find TDU show that
it only occurs when the stellar core-mass is greater than
about 0.6 M$_{\odot}$ (e.g., \citealt{1988ApJ...328..632B,
1989ApJ...344L..25L,1997ApJ...478..332S}),
but there is disagreement of
about $10\%$ on the exact value of this
minimum core-mass (see, e.g., \citealt{2003ApJ...586.1305L}).
The luminosity and surface abundances of carbon and $s$-process
elements in TP-AGB stars of mass $M\lesssim2$ M$_{\odot}$
depend strongly on this minimum core-mass. We have modelled the TDU
by using the fit to the models of
\citet{2002PASA...19..515K} from \citet{2004MNRAS.350..407I},
which includes the free parameter
$\Delta M_{\rm c}^{\rm min}$.
This parameter offsets the
minimum core-mass by a fixed amount compared to the
results of \citet{2002PASA...19..515K}, allowing stars of low
mass to undergo TDU even if they do not do so it in the detailed models.

\subsection{Third dredge-up efficiency}
Detailed evolution models disagree on the amount of third
dredge-up that occurs in a star of a particular mass and metallicity,
i.e., how deep the convective envelope penetrates the intershell
(see, e.g., \citealt{1996ApJ...473..383F} and
\citealt{1999A&A...344..617M} for discussion).
The amount of dredge-up can be measured by the
{\it third dredge-up efficiency}
\begin{equation}
\lambda=\frac{\Delta M_{\rm TDU}}{\Delta M_{\rm H}},
\end{equation}
where $\Delta M_{\rm TDU}$ is the reduction in mass of the
hydrogen exhausted core as a result of dredge-up and
$\Delta M_{\rm H}$ is the amount of mass by which the
hydrogen-exhausted core increased due to hydrogen burning
during the previous inter-pulse period. Thus, if $\lambda=0$
there is no dredge-up at all, while if 
$\lambda=1$ there is no net growth of the hydrogen-exhausted
core.
We apply the parametrization from
\citet{2004MNRAS.350..407I} in which
$\lambda$ grows asymptotically with pulse number to a maximum
value
\begin{equation}
\lambda_{\rm max}=\max\left(\lambda_{\rm min},
\lambda_{\rm max}^{\rm fit}\right).
\end{equation}
Here $\lambda_{\rm max}^{\rm fit}$ is a function of mass
and metallicity which fits the $\lambda_{\rm max}$ value of
the detailed models \citep{2002PASA...19..515K}
and the free parameter $\lambda_{\rm min}$ is used to set
a lower limit. In this way stars which show negligible TDU
efficiency in the detailed models are allowed to have a substantial 
TDU efficiency when $\lambda_{\rm min}$ is set to be
larger than zero. 

\subsection{Effectiveness of $^{13}$C as a neutron source}
\label{c13eff}
At the end of a TDU episode
protons from the bottom of the H-rich
envelope may penetrate beyond the chemical boundary into the
helium- and carbon-rich intershell. They react with $^{12}$C to produce
$^{13}$N, which decays into $^{13}$C forming the $^{13}$C pocket.
If there are enough protons, $^{13}$C will also capture protons and
produce $^{14}$N.
The $^{13}$C in the pocket is a
source of neutrons by means of
the ${\rm ^{13}C}(\alpha,n)^{16}{\rm O}$ reaction, but it has to
compete with $^{14}$N as a sink of free neutrons due to 
the $^{14}{\rm N}(n,p)^{14}{\rm C}$ reaction, which has a relatively
high neutron capture cross section.
As a result of this competition an {\it effective local
${\rm ^{13}C}$ abundance} can
be defined, which parametrizes the number of free neutrons per heavy
seed nucleus in
the intershell available for the $s$-process to take place.
The abundance ratios of
$s$-process elements in the ${\rm ^{13}C}$ pocket are
determined by this local $^{13}$C
abundance, which is scaled relative to a standard case.
This standard case was defined by \citet{1998ApJ...497..388G}
and \citet{2001ApJ...557..802B}, based on the
$^{13}$C pocket
found by \citet{1988ApJ...333L..25H} in their detailed evolutionary
calculations.
\citet{1998ApJ...497..388G} showed that with this pocket
low mass AGB stars of half solar metallicity
produce the main $s$-process component for the Sun.
We denote this scaled
effective $^{13}$C abundance as ${\rm ^{13}C}_{\rm eff}$,
so that the standard
case corresponds to ${\rm ^{13}C}_{\rm eff}=1$.

Comparing the absolute elemental abundances produced by
$s$-process models directly to observations is difficult
because they depend on the dilution of material from the intershell
into the envelope, which in turn depends on several uncertain factors
such as the size in mass of the ${\rm ^{13}C}$ pocket, the amount of TDU,
mass loss and mass accretion.
However, the $s$-process element abundance ratios remain mostly
unaffected by these processes, so they provide direct constraints
on ${\rm ^{13}C}_{\rm eff}$ (see $\S$\ref{results}).

\subsection{Size of the $^{13}$C pocket}
In order to calculate the
abundances in the stellar envelope we must know
the element mass fraction in the intershell at the moment when
the star undergoes TDU.
For $s$-process elements, we calculate these intershell abundances
by using the results of detailed models from
\citet{1998ApJ...497..388G}, which consist of a grid of
intershell elemental abundances at the time of TDU
for stellar masses of 1.5, 3 and 5 M$_{\odot}$,
metallicities $10^{-4}\leq Z\leq 0.02$, effective local ${\rm ^{13}C}$
abundance ($1/24\leq{\rm ^{13}C}_{\rm eff}\leq 2$) and
number of thermal pulse followed by TDU ($2\leq N_{\rm TDU}\leq 29$).
We know the mass of the $^{13}$C pocket and
of the intershell in these detailed models, which  
we use to trace back the local element mass fractions in the
$^{13}$C pocket just before the thermal pulse occurs.
We map a grid of $s$-process element
mass fractions in the $^{13}$C pocket for different stellar masses,
metallicities, ${\rm ^{13}C}_{\rm eff}$ and pulse number,
which is independent of the choice of
the size of the $^{13}$C pocket and that of the intershell.
We interpolate linearly on this grid and if the stellar mass
and/or pulse number needed for the calculation is out of the
grid bounds we use the result of the closest point on the grid.

Note that with this procedure we do not properly account for
the $^{22}$Ne neutron source. However, this neutron source does not
contribute to defining the overall $s$-process distribution, at
least for stellar models of mass $M\lesssim4$ M$_{\odot}$, which
are believed to be the counterparts of the observed $s$-process
enhanced stars that we will discuss in $\S$\ref{results}
(e.g., \citealt{2001ApJ...559.1117A}).

We assume that the $^{13}$C pocket mass is
a fixed fraction, $f_{\rm ^{13}C,IS}$, of the
mass of the intershell
at the moment just before TDU occurs, $M_{\rm IS}$,
which in turn is well approximated by the maximum mass
of the convective intershell pocket during the thermal pulse.
We employ the relation proposed by
\citet{1977ApJ...217..788I} for the maximum convective
intershell mass,
\begin{equation}
\log_{10}(M_{\rm IS})=-1.835+1.73M_{\rm c}-2.67M_{\rm c}^2,
\end{equation}
where $M_{\rm c}$ is the mass of the H-exhausted core.
This relation is derived for core
masses larger than 0.95M$_{\odot}$, but when extrapolated to
lower core masses it yields similar intershell mass values to those
considered by
\citet{1998ApJ...497..388G} in their detailed models.
The value of $f_{\rm ^{13}C,IS}$ can be constrained by
observations of surface abundances of $s$-process elements,
although these abundances also depend on the amount of dredge-up that
the star has experienced. This degeneracy is broken by
studying the abundances of other elements that are not synthesized
via the $s$-process, as discussed in $\S$\ref{lamb_pock}.

\subsection{Mass loss}
Mass loss is an uncertain factor in AGB evolution which
affects the enhancement of
$s$-process elements on the surface of a TP-AGB star
\citep{2003PASA...20..389S}. The mass loss rate determines
the stellar lifetime as a TP-AGB star by truncating it
when the envelope is lost.
In the models of \citet{1998ApJ...497..388G} the intershell
$s$-process abundance ratios vary from one pulse to another,
until after roughly 20 pulses
they reach an asymptotic value. The asymptotic values are
usually considered to be
the $s$-element ratios of a typical pulse and are often compared
to the observed ratios (e.g., \citealp{2001ApJ...557..802B,
2003PASA...20..389S,2004A&A...417..269R}).
Depending on the mass loss rate applied in the models,
a star with mass 
$M\lesssim1.5$M$_{\odot}$ can undergo from very
few thermal pulses (or none) to several tens, which affects its
surface composition and yield during the TP-AGB phase.
In this work we use the mass-loss prescription by
\cite{1993ApJ...413..641V}:
\begin{equation}
{\rm log}_{10}\frac{\dot{M}_{\rm AGB}}{{\rm M_{\odot}~yr^{-1}}}=
-11.4+0.025\left[\frac{P}{\rm day}-100
\max\left(\frac{M-2.5M_{\odot}}{\rm M_{\odot}},0\right)\right], 
\end{equation}
where log $P/$day $= -2.07+1.94$ log $R/$R$_{\odot}-0.9$ log
$M/$M$_{\odot}$ is the logarithm of the Mira pulsation period.
We do not attempt to vary the AGB mass-loss rate, but we discuss
the consequences of this assumption on our results in
$\S$\ref{disc}.

The mass-loss rate during the giant
branch phase affects the envelope mass at the moment
when a star begins its TP-AGB phase.
We have modelled mass-loss from giant branch stars with the
widely used prescription by \citet{1975psae.book..229R},

\begin{equation}  \label{eq:Reimers}
\dot{M}_{\rm GB}=4\times10^{-13}\eta_{\rm GB}
\left({L}/{\rm L_{\odot}}\right)\left({R}/{\rm R_{\odot}}\right)
\left({M}/{\rm M_{\odot}}\right)^{-1}{\rm M_{\odot}~yr^{-1}},
\end{equation}
with $\eta_{\rm GB}=0.3$, which is within the limits set by the
morphology of horizontal branch stars in the galactic globular
clusters \citep{1983ARA&A..21..271I,1996A&A...305..849C}.

   \begin{figure}
   \centering
   \includegraphics[width=0.5\textwidth]{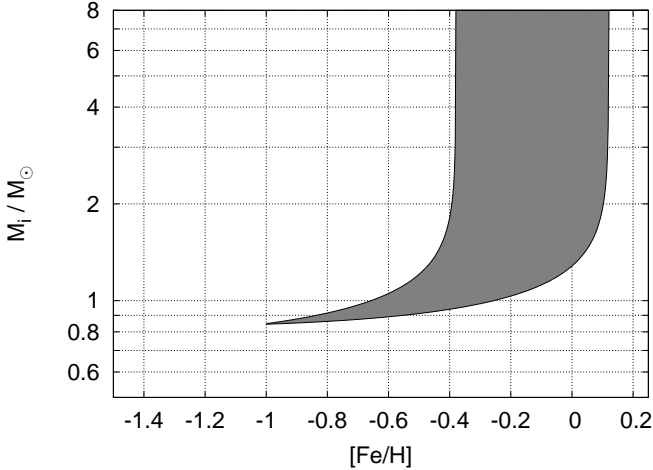}
      \caption{Range of initial masses, $M_{\rm i}$, of our population
	synthesis
	grid as a function of metallicity, depicted by the shaded area.
	For a given metallicity this shows the initial-mass range
	of stars which can be observed in the TP-AGB phase,
	with a maximum mass of 8 M$_{\odot}$.}
      \label{fig_AMR}
   \end{figure}

\section{Population synthesis}
\label{psyn}

We make use of our synthetic evolutionary models to
carry out population synthesis of single stars, on a grid
of 50 logarithmically spaced metallicities ranging
between $10^{-4}\leq Z\leq 0.025$
and 500 logarithmically spaced initial
masses, whose range depends on the metallicity and on whether we are
modelling a population of intrinsic or extrinsic $s$-enhanced stars.

In the case of intrinsic stars, for each metallicity we
apply the initial mass ranges shown in Fig. \ref{fig_AMR}, with 8
M$_{\odot}$ as a maximum.
We derive these ranges from combining our stellar evolution models
with the age$-$metallicity relation (AMR)
for the galactic disk of \citet{2004MNRAS.351..487P}
\begin{equation}
\label{amreq}
\left[\frac{\rm Fe}{\rm H}\right]=
{\rm log_{10}}\left(\frac{t-17.82}{-23.24}\right)\pm 0.25{\rm ~dex},
\end{equation}
where the notation for the element number density ratio
${\rm \Big[\frac{X}{Y}\Big]=
\log_{10}\left(\frac{n_X}{n_Y}/\frac{n_{X_{\odot}}}{n_{Y_{\odot}}}\right)}$
is used
and $t$ is the age in Gyr, with the age of the
Universe (13.7 Gyr) as its upper limit.
Then, from our single-star models we consider the range of initial
masses for which a star spends at least some of its TP-AGB life
within the age range allowed by the AMR. The probability for an intrinsic
star to have a given abundance is weighted
by the time it spends showing
this abundance and by its contribution according to the initial mass
function (IMF). We use the IMF, $\xi$, of \citet{1993MNRAS.262..545K},
given by
\begin{equation}
\xi(M_{\rm i})=
\begin{cases}
  0.035M_{\rm i}^{-1.3}& \text{if 0.08 $<M_{\rm i}<$ 0.5},\\
  0.019M_{\rm i}^{-2.2}& \text{if 0.5 $<M_{\rm i}<$ 1.0},\\
  0.019M_{\rm i}^{-2.7}& \text{if 1.0 $<M_{\rm i}$},\\
\end{cases}
\end{equation}
where $M_{\rm i}$ is the initial stellar mass in solar units.

Given that
we employ single stellar models, we
treat the case of extrinsic stars indirectly.
These stars have acquired their $s$-process enhancements by accreting
part of the mass ejected by an initially more massive binary companion.
Therefore we assume a simple model in which the 
initial mass of the companion is assumed to be distributed according 
to the single-star IMF. Hence the
probability of an extrinsic star to show a given elemental
abundance is weighted by the amount of mass which is lost by its
companion in the form of that element
and by the IMF weight of this companion. 
Because mass transfer could have occurred at any time since the
formation of the system, we consider all
stars whose evolution is beyond the TP-AGB phase,
which in each metallicity bin corresponds to initial masses greater
than the lower limit in Fig. \ref{fig_AMR} and smaller than
8 M$_{\odot}$. For the models with [Fe/H] $<-1$ we consider
0.8 M$_{\odot}$ to be the lower initial mass limit.

\section{Results}
\label{results}
In this section we present the results of our population synthesis
models and compare them to different sets of observational data.
The observations provide measurements of the surface abundances of
$s$-process elements relative to that of iron. Following the
convention, we calculate for each star in our
grid its {\it heavy} $s$-process (hs) element abundance ratio,
\begin{equation}
\rm{
\left[\frac{hs}{Fe}\right]=\frac{1}{5}\left(\left[\frac{Ba}{Fe}\right]+
\left[\frac{La}{Fe}\right]+\left[\frac{Ce}{Fe}\right]+
\left[\frac{Nd}{Fe}\right]+\left[\frac{Sm}{Fe}\right]\right)},
\end{equation}
and {\it light} $s$-process (ls) element abundance ratio,
\begin{equation}
\rm{
\left[\frac{ls}{Fe}\right]=\frac{1}{2}\left(\left[\frac{Y}{Fe}\right]+
\left[\frac{Zr}{Fe}\right]\right)}.
\end{equation}
These ratios
are altered by dilution in the stellar envelope,
which is affected by
several uncertain factors
such as the size in mass of the ${\rm ^{13}C}$ pocket,
the amount of TDU, mass loss and,
in the case of binaries, mass accretion. We also study the
ratio of hs-elements to ls-elements,
\begin{equation}
\rm{\left[\frac{hs}{ls}\right]=\left[\frac{hs}{Fe}\right]-
\left[\frac{ls}{Fe}\right]}.
\end{equation}
This ratio reaches asymptotically the [hs/ls] values of the intershell
after a number of pulses, making it mostly unaffected by dilution.

\begin{figure}
   \centering
   \includegraphics[width=0.5\textwidth]{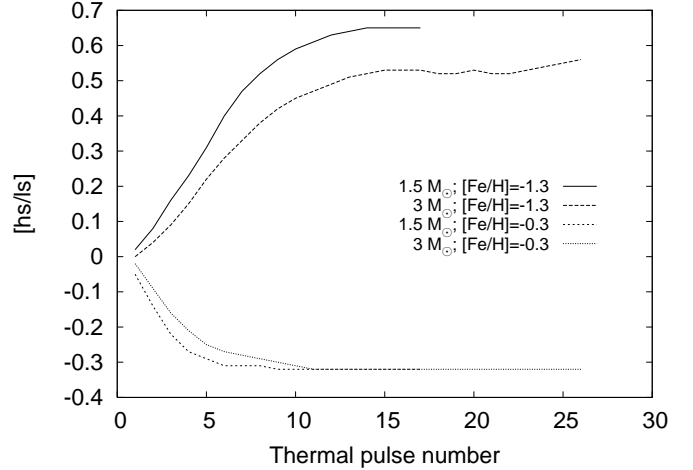}
      \caption{Envelope [hs/ls] during the evolution of 1.5 M$_{\sun}$
	and 3 M$_{\sun}$ star models of different metallicity,
	as indicated in the panel, and with $^{13}$C$_{\rm eff}$=1.}
      \label{Fig:hsls}
   \end{figure}

The asymptotic envelope ratio is therefore only determined by 
${\rm ^{13}C}_{\rm eff}$ (see $\S$\ref{c13eff}), but not by
other free parameters in our model, such as the size of the
${\rm ^{13}C}$ pocket ($f_{\rm ^{13}C,IS}$).
Fig. \ref{Fig:hsls} shows envelope [hs/ls] ratios as obtained from the
models of \citet{2001ApJ...557..802B}, which in each model reach their
corresponding intershell asymptotic value,
but take a number of pulses to do so.
Hence, considering stars
that have experienced only few TDU episodes, either because they
are in an early stage of the TP-AGB phase or because they have masses
$M\lesssim1.5$ M$_{\odot}$,
results in a natural spread of [hs/ls] values.
Only in AGB stars with $M\gtrsim5$ M$_{\odot}$ is the
amount of dredged-up material is so small that
the surface [hs/ls] ratios differ from those of the
intershell.

To compare these abundance ratios with the observed data
we have binned the resulting number distribution with a
resolution of 0.1 dex. Whenever a star has a certain
abundance ratio, its probability is added to the
corresponding bin. This is done separately
for the different metallicity bins and in 
each we normalize the resulting 
probability distribution by dividing it by its highest
value.

   \begin{figure*}
   \centering
   \includegraphics[width=\textwidth]{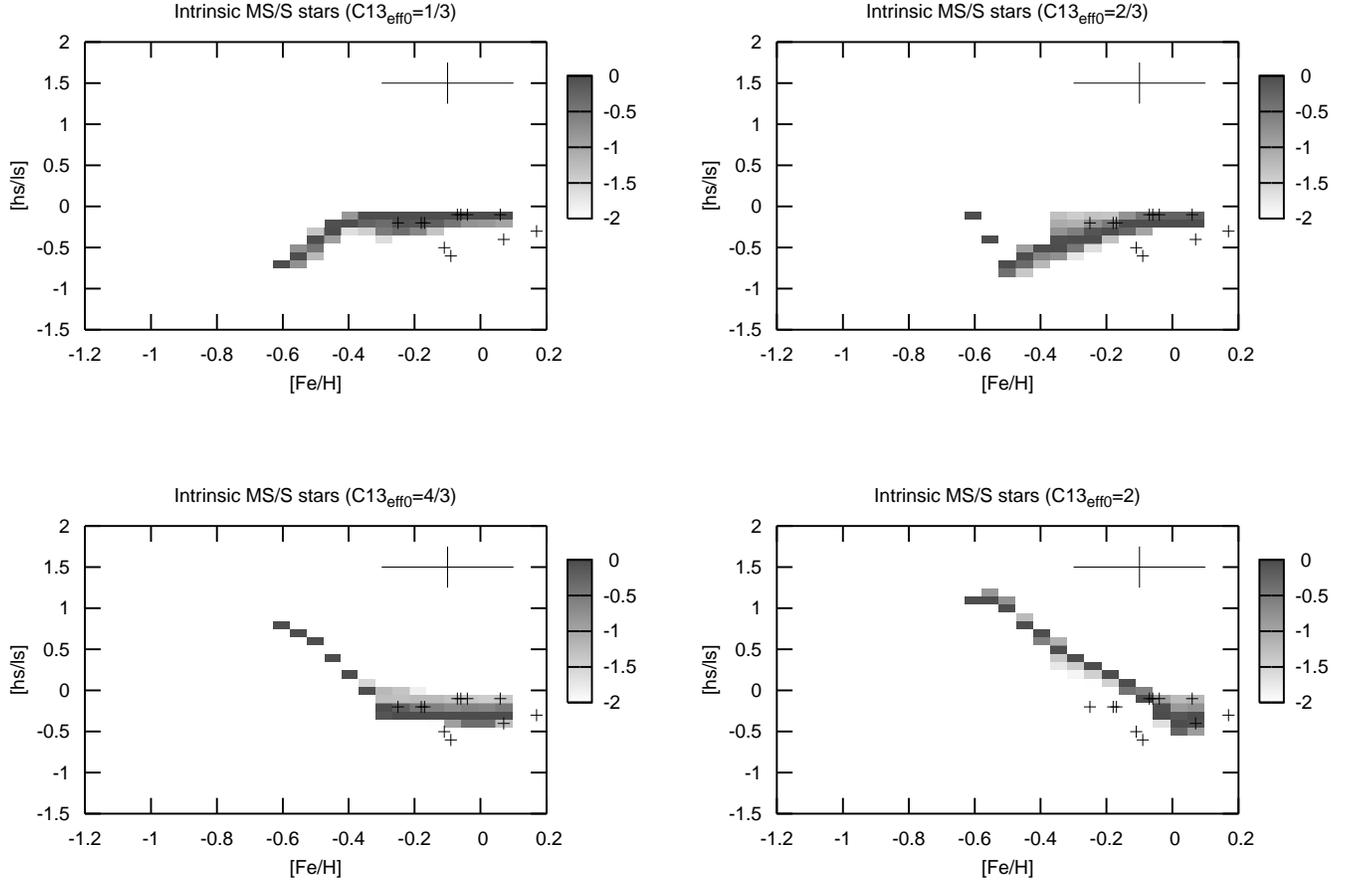}
      \caption{Intrinsic MS/S stellar population synthesis
	results	compared to the observations.
	The panels show our results calculated
	with four increasing $^{13}{\rm C}_{\rm eff}$ values
	as indicated above the panels.
	The grey scale is a logarithmic
	measure of the normalized number distribution of stars over
	[hs/ls]. The crosses are the
	observational data gathered by \citet{2001ApJ...557..802B},
	which have an average
	error given by the size of the cross in the
	upper right of each plot.}
      \label{figpsyn_M}
   \end{figure*}

We select to be $s$-enhanced those stars which have
[ls/Fe] and/or [hs/Fe] larger than 0.1 dex. 
First we analyze the [hs/ls] ratios of
intrinsic $s$-enhanced stars, which have a strong dependence only
on the $^{13}{\rm C}_{\rm eff}$ value. In all the runs we use
$\Delta M_{\rm c}^{\rm min}= -0.065$M$_{\odot}$,
$\lambda_{\rm min}=0.2$ and $f_{\rm ^{13}C,IS}=1/40$,
based on our attempts to fit the post-AGB star observations
of [Zr/Fe] ratios and carbon abundances, 
as explained in detail in $\S$\ref{lamb_pock}.
In $\S$\ref{extrinsic} we study the [hs/ls]
ratios, but this time of extrinsic $s$-enhanced stars, which give
clues on $^{13}{\rm C}_{\rm eff}$ at metallicities
lower than that of the galactic disk.

   \begin{figure*}
   \centering
   \includegraphics[width=\textwidth]{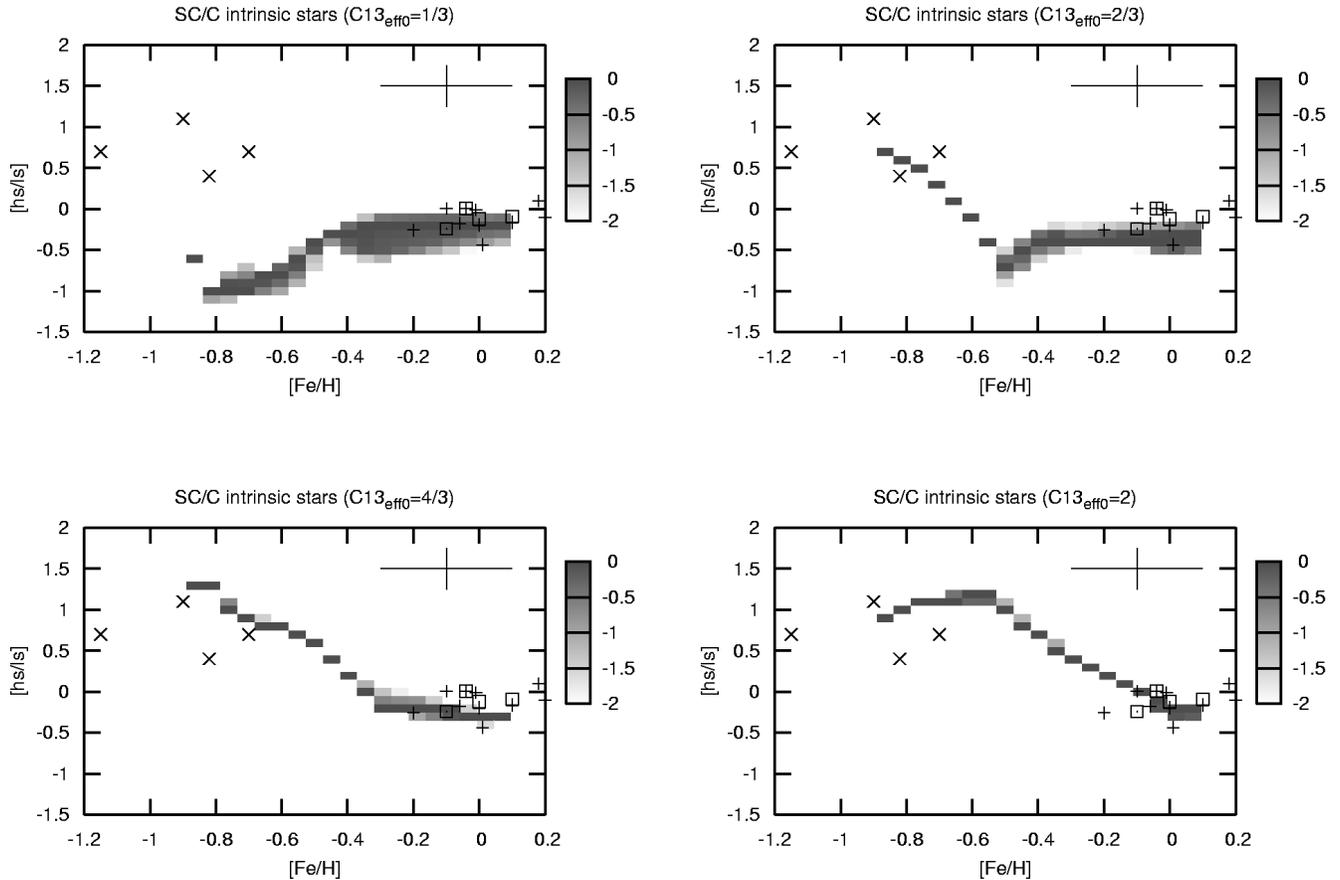}
      \caption{As Fig. \ref{figpsyn_M}, but for intrinsic SC
	and C stars. The horizontal crosses are observational data
	of intrinsic SC/C star gathered by \citet{2001ApJ...557..802B},
	complemented with data of \citet{2002ApJ...579..817A}.
	The squares are SC/C stars in which the presence of Tc is
	doubtful \citep{2002ApJ...579..817A} and
	the diagonal crosses are halo C stars from
	\citet{2001ApJ...557..802B}.
	The observations have an average error given by the size
	of the cross in the upper right of each plot.
              }
         \label{figpsyn_SC}
   \end{figure*}

\subsection{Intrinsic $s$-enhanced stars}
\label{intrinsic}
\subsubsection{[hs/ls] ratios of MS/S stars}
\label{MS_S}
MS/S stars are late-type giants that show lines
of ZrO and TiO, indicating that they are enhanced in $s$-element
abundance, have low surface temperature and their C/O ratio is
smaller than unity.
To account for these stars we
select from our population synthesis
$s$-enhanced stars in the TP-AGB phase of evolution with a
surface effective temperature less than 3500 K and
C/O $<1$ in their envelope.
We compare our results for the [hs/ls] ratio
to observational data of intrinsic MS/S stars gathered
by \citet{2001ApJ...557..802B}.
In Fig. \ref{figpsyn_M} our results
only appear at [Fe/H] $\gtrsim-0.6$ because
low-metallicity stars that under-go TDU episodes
are too hot to be classified as MS/S objects.
This is consistent with the observations.
Increasing $^{13}{\rm C}_{\rm eff}$ in our
models increases the [hs/ls] ratios on the
low-metallicity side of the plot, but at
high metallicities the [hs/ls] ratios are decreased.
This is explained as follows: If the number of heavy iron seeds
is decreased for a given number of
available free neutrons, then the [ls/Fe] ratio
increases due to $s$-process reactions and [hs/ls]
decreases, as is observed at high metallicities in our results.
But if [Fe/H] is decreased even more then the ls-element synthesis
saturates and the hs-element abundance increases with respect to iron,
raising both ratios [hs/Fe] and [hs/ls], as can be seen in our models
at low metallicities \citep{1999ApJ...521..691T,2001ApJ...557..802B}.
Thus $^{13}{\rm C}_{\rm eff}$ basically
determines at which metallicity our models will show [hs/ls]
ratios below solar and where they will rise to solar and beyond. This
effect cannot be seen clearly in our MS/S star synthesis due to the
lack of $s$-enhanced MS/S stars below [Fe/H] $\sim-0.6$,
but is common to all the stellar populations that we
have synthesized and is more evident in SC and C stars and post-AGB
stars ($\S$\ref{SCandC} and $\S$\ref{postagb}, respectively).
In contrast with other studies where only the asymptotic values of
[hs/ls] are considered (e.g., \citealt{2000A&A...362..599G,
2001ApJ...557..802B}), a natural
spread in this ratio is observed in our results. This is
explained by our tracing of the whole TP-AGB evolution, where the
initial dozen pulses have different $s$-process ratios.
Stars of mass $M\lesssim1.5$ M$_{\odot}$,
which are the most numerous due to the IMF and
which do not undergo more than 10$-$20 pulses, contribute most to this
effect.
Fig. \ref{figpsyn_M} shows that the best fit to the observed
distribution of MS/S stars is obtained with
$^{13}{\rm C}_{\rm eff}\approx4/3$ and a spread in
$^{13}{\rm C}_{\rm eff}$ values is not needed to match all the
observations. Given that the error bars on the individual
observations are large (as indicated in Fig. \ref{figpsyn_M}),
and that the observed [hs/ls] values
are close to 0, the value of $^{13}{\rm C}_{\rm eff}$ is not
strongly constrained by these observations.
The (on average)
negative values of [hs/ls] are matched, within observational errors,
by the models if $^{13}{\rm C}_{\rm eff}$ is between 2/3 and 2.
As the [hs/ls] ratio is not sensitive to
dilution, changes in the other free parameters do not affect
these results. 

   \begin{figure*}
   \centering
   \includegraphics[width=\textwidth]{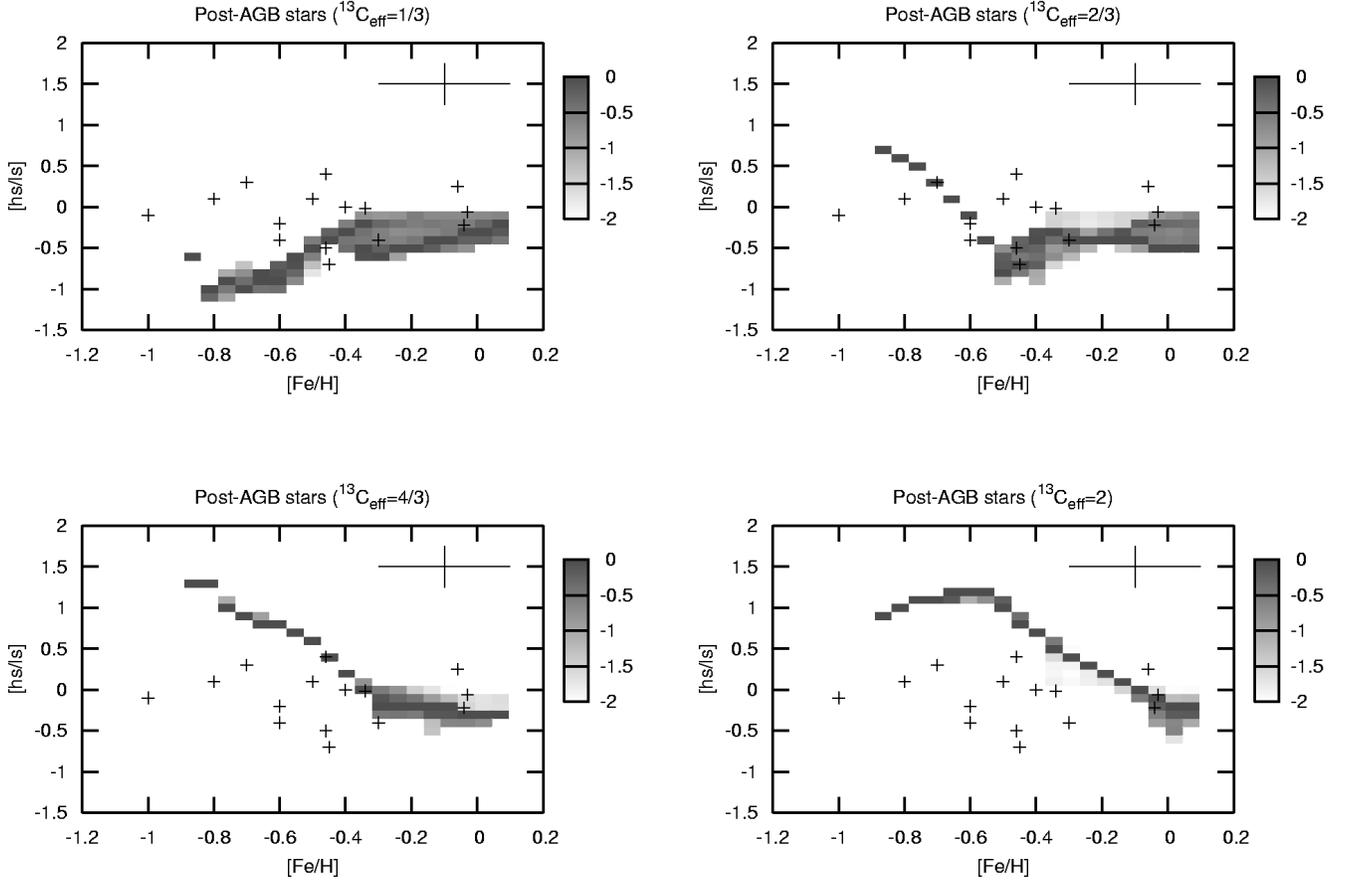}
      \caption{As Fig. \ref{figpsyn_M}, but for post-AGB stars
	The horizontal crosses are the
	observational data (see references in the text), which
	have an average error given by the size of the
	cross in the upper right of each plot.}
      \label{figpsyn_post}
   \end{figure*}

\subsubsection{[hs/ls] ratios of SC and C stars}
\label{SCandC}
SC and C stars are late-type giants that show lines of
carbide molecules in their spectrum, which indicates that
there is more C than O by number. In the case of the SC stars,
the C/O ratio is close to unity, given that they also show
ZrO lines. To reproduce these stars we select from our
population synthesis results those $s$-enhanced TP-AGB stars
which have C/O $>1$. We compare our synthetic
[hs/ls] ratios with the observational data gathered by
\citet{2001ApJ...557..802B} and from
\citet{2002ApJ...579..817A} as shown in Fig. \ref{figpsyn_SC}. 
The spread in the [hs/ls] ratio is explained in the same way
as for MS/S stars. The spread is smaller at
low metallicities due to the small range of initial stellar
masses considered at these metallicities resulting from the
AMR (see Fig. \ref{fig_AMR}).
The effect of changing
$^{13}{\rm C}_{\rm eff}$ on our synthetic [hs/ls] ratios is 
clearer in this population of stars than in the MS/S stars,
particularly for ${\rm [Fe/H]}\lesssim-0.4$.
As in the case of the MS/S stars, the other free parameters do
not affect the [hs/ls] ratios.
Most of the observed SC and C stars cluster around
${\rm [Fe/H]}\approx0$ and slightly negative values of [hs/ls].
These stars can be fitted with any value of $^{13}{\rm C}_{\rm eff}$.
The diagonal crosses in Fig. \ref{figpsyn_SC} are
halo C stars suspected to be intrinsic.
They are not expected to follow the same AMR as
those in the galactic disk, but the fact that they are old and
of low mass is approximately modelled by our AMR. Three of these
stars are matched by our models for values of
$^{13}{\rm C}_{\rm eff}\gtrsim2/3$, but a lower
$^{13}{\rm C}_{\rm eff}$ cannot
reproduce their [hs/ls].
The lack of stars below [Fe/H] $\sim-1$ in our models is due to
the fact that, with our adopted AMR,
lower metallicity stars do not become TP-AGB stars
within the age of the Universe. This effect is observed in
all our results involving intrinsic stars and is consistent
with the observations.

\subsubsection{[hs/ls] ratios of post-AGB stars}
\label{postagb}
Post-AGB stars are in the fast
evolutionary
phase between the AGB and white dwarf track. They suffer
from strong mass loss and are hot, but not hot enough to
ionize their circumstellar medium.
We select them from our models by choosing those $s$-enhanced
TP-AGB stars that have less than 0.03M$_{\odot}$ left in their
envelopes. We compare our synthetic population to the
observational data gathered by \citet{2001ApJ...557..802B},
\citet{2003ARA&A..41..391V},
\citet{2004A&A...417..269R} and \citet{2005A&A...443..297G}. 
Fig. \ref{figpsyn_post} shows a spread, mostly at high metallicities,
which is due to the different initial stellar masses taken into account
in our population synthesis, the range of which becomes narrower towards
low metallicities as we follow the AMR.
Stars with $M\lesssim1.5$ M$_{\odot}$ experience a small number of
TDU episodes.
After their last thermal pulse (when they are observed as post-AGB
stars) they have not yet reached the asymptotic [hs/ls] ratio,
which contributes to the spread observed.
Most of the data can be fitted with
$2/3\lesssim~^{13}{\rm C}_{\rm eff}\lesssim4/3$, within the
observational errors. There is only one exception at [Fe/H]$\sim-1$,
which apparently needs a somewhat smaller $^{13}{\rm C}_{\rm eff}$.
This object, IRAS07134+1005, has a very high heliocentric 
velocity \citep{2000A&A...354..135V} suggesting
that it belongs to the galactic halo rather than the galactic disk. 

\subsubsection{Zr and C abundances of post-AGB stars}
\label{lamb_pock}
\begin{figure}
   \centering
   \includegraphics[width=0.5\textwidth]{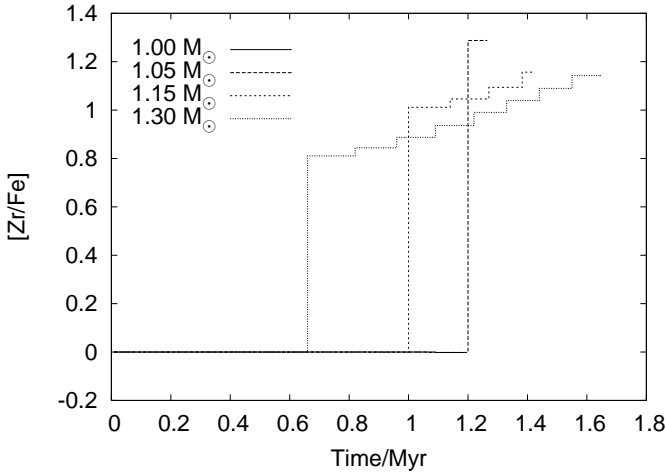}
      \caption{Enhancement of Zr in TP-AGB stars as function
	of time. The models shown are for [Fe/H] $=-0.5$, with
	initial masses as indicated in the figure.
	When the stars reach the post-AGB phase, i.e., by the
	end of their evolution, there is a strong Zr
	enhancement ($\sim1$ dex) as long as at least one
	$s$-enriched dredge-up episode has taken place. If no
	dredge-up occurs there is no enhancement at all, as
	seen in the case of the 1M$_{\odot}$ star.}
      \label{pagbenh}
   \end{figure}

Post-AGB stars can also be used to provide constraints
on the minimum core-mass for TDU, the TDU efficiency and the
size of the $^{13}$C pocket.
The observed values of [Zr/Fe] show a
bimodal distribution: ${\rm [Zr/Fe]}\approx0$ and
$1\lesssim{\rm [Zr/Fe]}\lesssim2$, which suggests that some post-AGB
stars did not experience TDU, while others did
and suffered a strong enhancement of Zr \citep{2003ARA&A..41..391V}.
This can be explained when we consider low mass stars
($M\lesssim1.5$ M$_{\odot}$)
at sub-solar metallicities. They have low-mass envelopes which
easily become strongly $s$-enhanced. As shown in Fig. \ref{pagbenh},
only one TDU episode after the intershell material has become
$s$-enriched is enough to raise their envelope abundance ratios by
almost 1 dex.
We compare the observational [Zr/Fe] ratios of
\citet{2003ARA&A..41..391V}, \citet{2004A&A...417..269R} and
\citet{2005A&A...443..297G} to our models, from which we
consider all post-AGB stars, both $s$-enhanced and not $s$-enhanced.

   \begin{figure}
   \centering
   \includegraphics[width=0.5\textwidth]{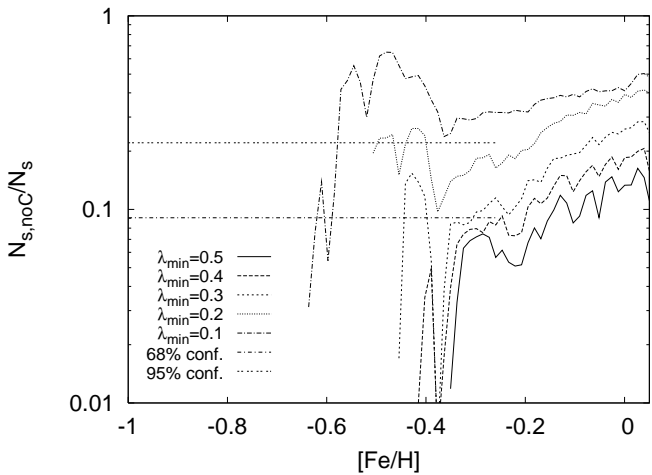}
      \caption{Ratio of the number of $s$-enhanced
	post-AGB stars which have ${\rm C/O}<1$ ($N_{s,{\rm noC}}$)
	to the number of $s$-enhanced post-AGB stars
	($N_{s}$) from our synthetic models. The
	curves show the ratios calculated with different values
	of $\lambda_{\rm min}$, as indicated in the figure.
	The straight dash-dotted and double-dotted lines show the
	upper limit to the $N_{s,{\rm noC}}/N_{s}$ ratio
	at [Fe/H] $\lesssim-0.3$,  with
	a confidence of 68\% and 95\%, respectively, given the
	distribution of the observed data.}
      \label{nonc_c_post}
   \end{figure}

Intrinsically $s$-enhanced post-AGB stars are observed at
metallicities as low as ${\rm [Fe/H]}\sim-1.0$
\citep{2003ARA&A..41..391V}.
When taking into account the AMR,
this implies that at metallicities [Fe/H] $\lesssim-0.7$
stars with initial masses above 0.9 M$_{\odot}$ must experience
TDU. At [Fe/H] $\lesssim-0.3$, the observed number of
$s$-enhanced stars (which experienced TDU) is comparable
to that of stars with ${\rm [Zr/Fe]}\approx0$ (which did not
undergo TDU).
Thus, taking into account
the IMF and the initial
mass range given by the AMR at low metallicities, we estimate
the minimum initial mass at which TDU must take place so that
the number of $s$-enhanced stars is similar to that of
non-enhanced stars.
Applying this in our models, we find the corresponding minimum
\emph{core} mass for TDU as a function of metallicity.
The resulting relation between minimum core mass and metallicity
is best modelled by the fitting formula of \citet{2002PASA...19..515K},
with an offset $\Delta M_{\rm c}^{\rm min}=-0.065$M$_{\odot}$.
This outcome depends somewhat on the
adopted age-metallicity relation which sets the initial mass range of
TP-AGB stars at each metallicity (Fig.~\ref{fig_AMR}), but the
sensitivity of the mass of a low-mass star to its lifetime is only
slight. Our result depends more strongly on the choice of mass
loss. To test this we apply different mass-loss rates on the TP-AGB
with the prescription of \citet{1975psae.book..229R},
Eq.~(\ref{eq:Reimers}), by varying the coefficient $\eta_{\rm GB}$
from 0.5 to 3.0, and find that $0.055\lesssim\Delta M_{\rm c}^{\rm
min}\lesssim0.075$.  This result is consistent with the findings of,
e.g., \citet{1993A&A...267..410G}, \citet{1999A&A...344..123M},
\citet{2004MNRAS.350..407I} and \citet{2005MNRAS.356L...1S} when
studying the carbon star luminosity functions of the Small and
Large Magellanic Cloud.

\begin{figure*}
   \centering
   \includegraphics[width=\textwidth]{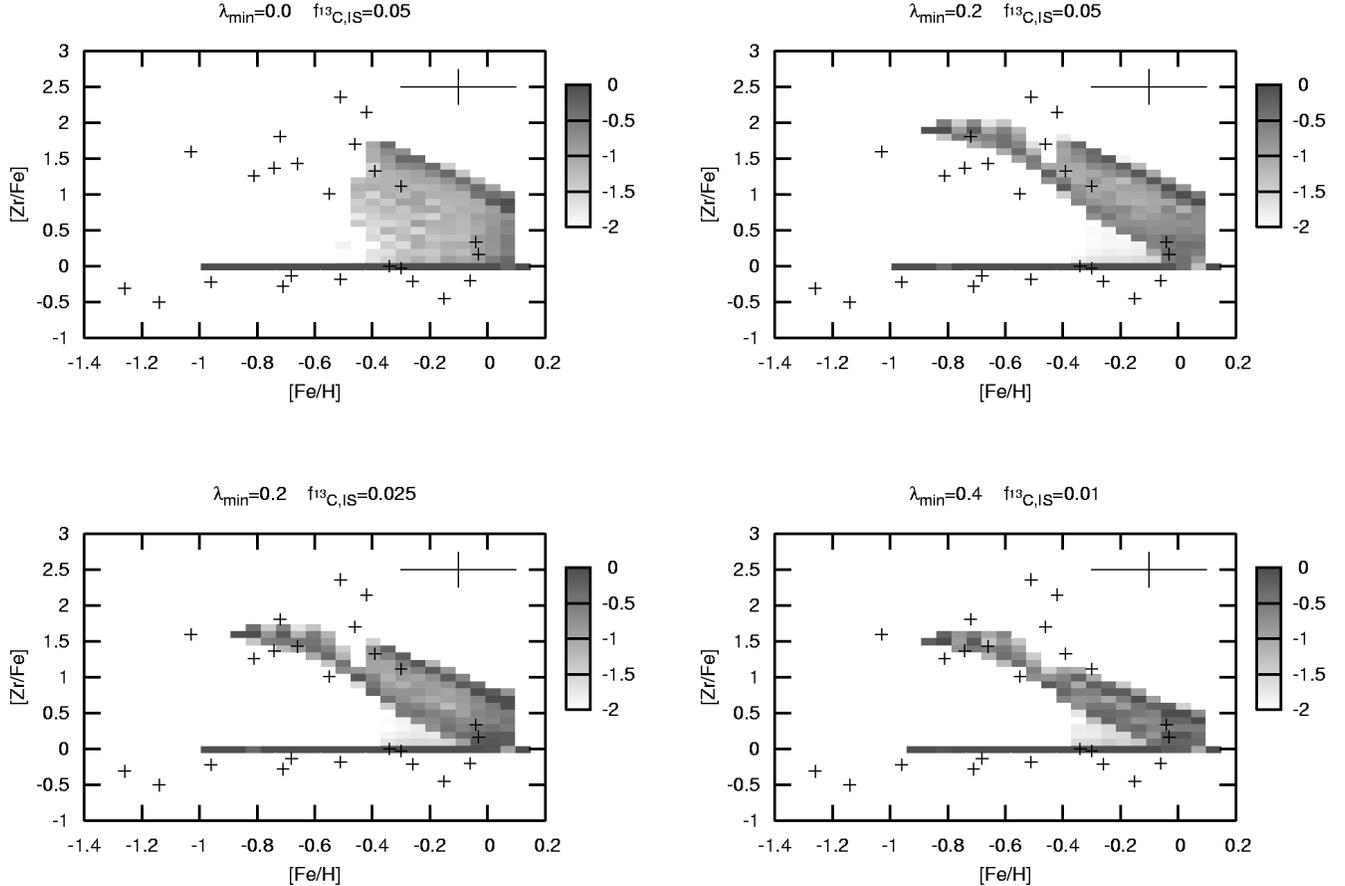}
      \caption{Post-AGB stellar population synthesis
	results	compared to the observations.
	The grey scale is a logarithmic
	measure of the normalized number distribution of stars in the
	[Zr/Fe]$-$[Fe/H] plane for $^{13}{\rm C}_{\rm eff}=2/3$.
	As indicated above the panels, plots are presented for different
	values of $\lambda_{\rm min}$ and $f_{{\rm ^{13}C,IS}}$.
	The crosses are the
	observational data (see references in the text), which
	have an average error given by the size of the
	cross in the upper right of each plot.}
      \label{figpsyn_post_Zr}
   \end{figure*}

\begin{figure*}
   \centering
   \includegraphics[width=\textwidth]{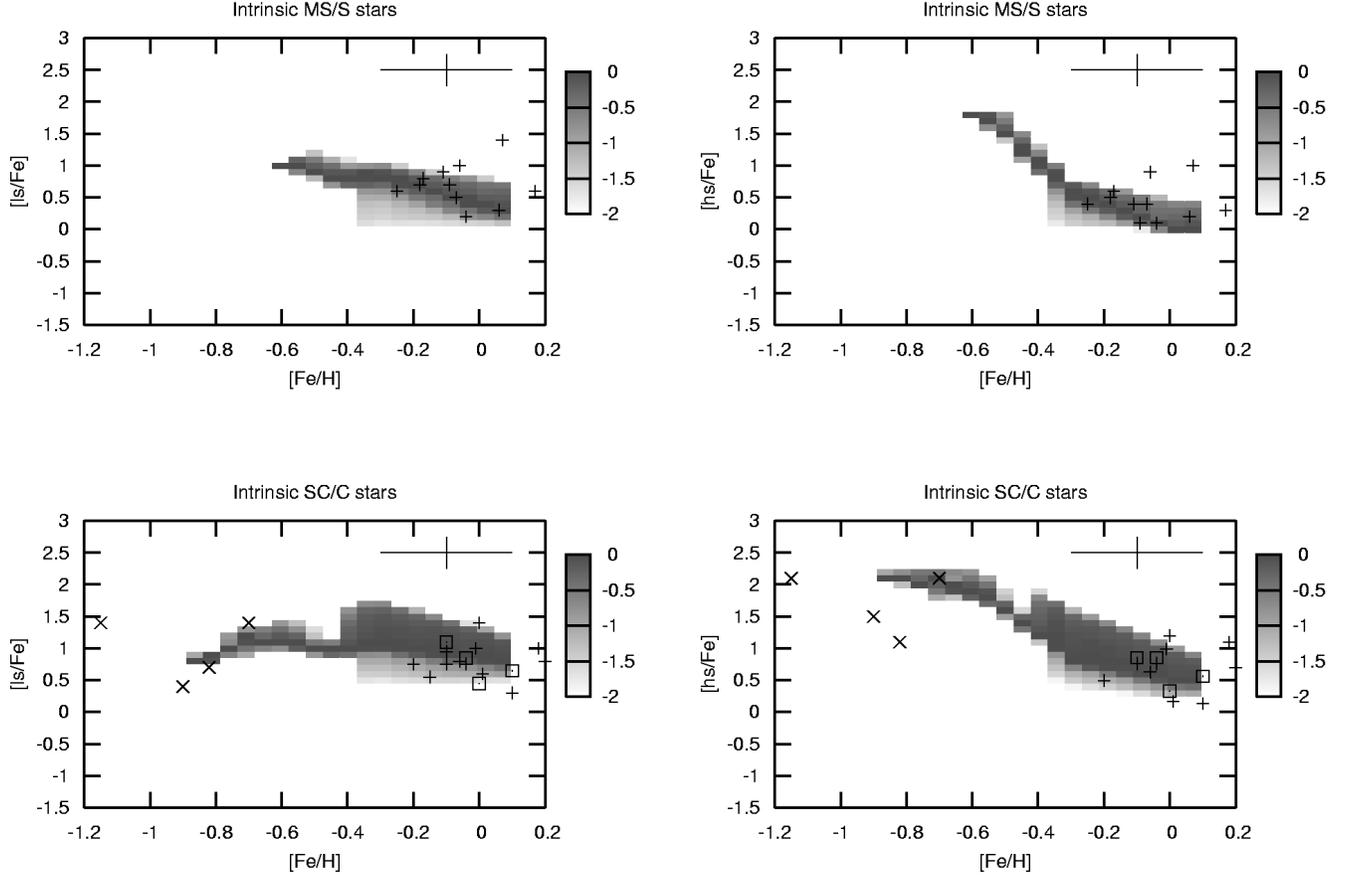}
      \caption{Stellar population synthesis
	results	of MS/S stars (top panels) and
	SC/C stars (bottom panels) compared to the
	observations.
	The grey scale is a logarithmic
	measure of the normalized number distribution of stars in the
	[ls/Fe]$-$[Fe/H] plane (left panels) and the
	[hs/Fe]$-$[Fe/H] plane (right panels), using
	$^{13}{\rm C}_{\rm eff}=4/3$ (as derived in
	$\S$\ref{MS_S} and $\S$\ref{SCandC}),
	$\lambda_{\rm min}=0.2$
	and $f_{{\rm ^{13}C,IS}}=0.025$.
	The observational data are from the same references
	and with the same coding as those of Fig. \ref{figpsyn_M}
	and Fig. \ref{figpsyn_SC}, which
	have an average error given by the size of the
	cross in the upper right of each plot.}
      \label{fig_MSC}
   \end{figure*}

We now address
the question how much intershell material is
dredged up to the surface. The surface
$s$-process enhancement in our models is controlled
by the dredge-up efficiency ($\lambda$) and
by the fractional mass of the $^{13}$C pocket relative
to the intershell ($f_{{\rm ^{13}C,IS}}$). The carbon
enrichment in the envelope depends only on the amount
of dredge-up, independent of the size of the $^{13}$C pocket.
The observations show that below [Fe/H] $\sim-0.3$ all
$s$-enhanced post-AGB stars have C/O $>1$ and that the
non-enhanced ones have C/O $<1$ \citep{2003ARA&A..41..391V}.
We use this information to
break the degeneracy between $\lambda$ and $f_{{\rm ^{13}C,IS}}$.
Let $N_{s{\rm ,noC}}$ be the observed number of
$s$-enriched post-AGB stars that have C/O $<1$ and $N_{s}$ be
total number of $s$-enriched post-AGB objects.
Fig. \ref{nonc_c_post} illustrates the ratio $N_{s{\rm ,noC}}/N_{s}$
as a function of metallicity calculated with our models for
different values of $\lambda_{\rm min}$. With a simple
application of Bayesian statistics we calculate that, given the
observed data distribution, $N_{s{\rm ,noC}}/N_{s}=0$ and $N_{s}=11$,
there is a 32\% probability that $N_{s{\rm ,noC}}/N_{s}>0.09$ and
only a 5\% probability that $N_{s{\rm ,noC}}/N_{s}>0.23$.
The latter sets the 95\% confidence limit $\lambda_{\rm min}\gtrsim0.2$,
as shown in Fig. \ref{nonc_c_post}.
With this lower limit we calibrate $f_{\rm ^{13}C,IS}$
to fit the Zr enhancement of post-AGB stars.
Fig. \ref{figpsyn_post_Zr} shows our population synthesis
results for the [Zr/Fe] ratio calculated with different values of
$\lambda_{\rm min}$ and $f_{\rm ^{13}C,IS}$, compared to
the observations. The upper left panel shows our results for
$\lambda_{\rm min}=0$. With this choice there are no $s$-enhanced
post-AGB stars at low metallicities (${\rm [Fe/H]}\lesssim-0.4$)
since these would all be low-mass stars due to the AMR and
they experience negligible dredge-up. This is in contradiction
to the observations.
Moreover, as shown above (Fig. \ref{nonc_c_post}),
a choice of $\lambda_{\rm min}<0.2$ is not compatible
with the number of carbon enhanced stars.
If we choose $\lambda_{\rm min}=0.2$ (upper right panel)
then low-metallicity $s$-enhanced post-AGB stars are indeed produced.
The large spread in the $s$-enhancement at higher metallicities
is due to stars with mass $M\gtrsim3$ M$_{\odot}$.
However, the choice $f_{\rm ^{13}C,IS}=0.05$, which roughly corresponds to
that used by \citet{1998ApJ...497..388G} for their detailed nucleosynthesis
calculations, gives too much Zr for many of the
$s$-enhanced post-AGB stars with ${\rm [Fe/H]}\lesssim-0.5$.
The lower right panel shows that most of the observations can
be reproduced with $\lambda_{\rm min}=0.2$ and
$f_{\rm ^{13}C,IS}\approx 1/40$, with two exceptions which are
extremely $s$-enhanced.
From the constraint that all $s$-enhanced post-AGB stars are carbon
stars, there is a 68\% confidence lower limit on $\lambda_{\rm min}$
of 0.4.
This implies that $f_{\rm ^{13}C,IS}\approx 1/100$ is needed to fit the
post-AGB observations (lower right panel of
Fig \ref{figpsyn_post_Zr}). However, this choice of $f_{\rm ^{13}C,IS}$
gives synthesized [ls/Fe] and [hs/Fe] ratios which are too small
to reproduce those observed in MS/S, SC and C stars.
Fig. \ref{fig_MSC} shows a comparison of the observed
[ls/Fe] and [hs/Fe] of MS/S, SC and C stars
to our models, calculated with
$\Delta M_{\rm c}^{\rm min}= -0.065$M$_{\odot}$,
$\lambda_{\rm min}=0.2$, $f_{\rm ^{13}C,IS}=1/40$
and $^{13}{\rm C}_{\rm eff}=4/3$.
Most of the MS/S star abundances
are fitted within the errors, with two high-metallicity
exceptions that are only matched with a larger choice of
$^{13}{\rm C}_{\rm eff}$ or $f_{\rm ^{13}C,IS}$.
The $s$-process abundances of most of the SC and C stars in the
galactic disk are also matched well using the same choice
of free parameters, except for three low metallicity
objects that need a smaller $^{13}{\rm C}_{\rm eff}$.

From the combined evidence of MS/S, SC, C and post-AGB stars,
we find that with a few exceptions, all observations of intrinsic
$s$-enhanced stars can be matched by models
with the following set of parameters:
\begin{itemize}
\item $\Delta M_{\rm c}^{\rm min}= -0.065$M$_{\odot}$
\item $\lambda_{\rm min}=0.2$
\item $f_{\rm ^{13}C,IS}=1/40$
\item $2/3\leq~^{13}{\rm C}_{\rm eff}\leq4/3$.
\end{itemize}
However, the
constraints we have set here depend somewhat on our choice of other rather
uncertain parameters, namely the mass-loss rate and
age$-$metallicity relation (see \S~\ref{disc}).

\subsection{Extrinsic $s$-enhanced stars}
\label{extrinsic}

   \begin{figure*}
   \centering
   \includegraphics[width=\textwidth]{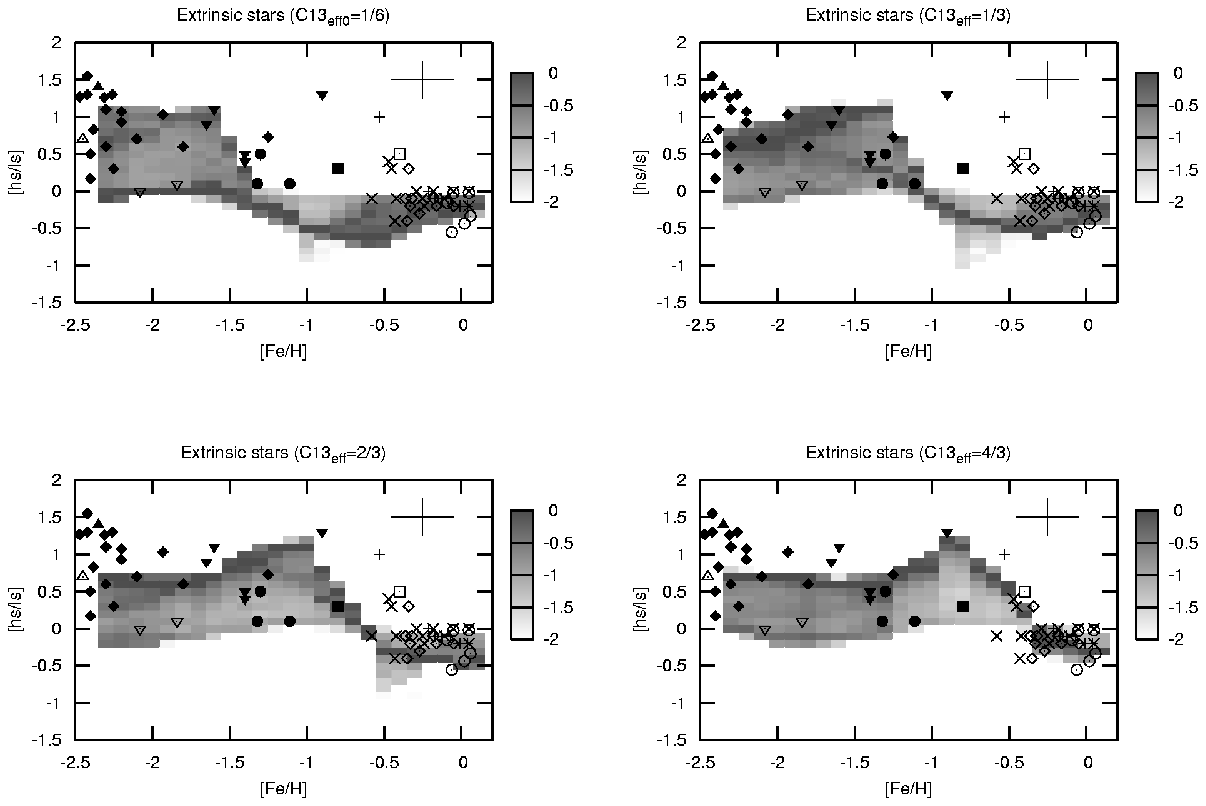}
      \caption{Population synthesis results for single-star yields
	compared to observations of extrinsic $s$-enhanced stars
	for different values of $^{13}$C$_{\rm eff}$.  
	The grey scale is a normalized logarithmic distribution of [hs/ls]
	for a population of stars as a function of metallicity.
	The symbols indicate observed data from different
	stellar types and references as follows:
	$+$ MS/S stars; $\times$
	Ba II giants; $\Diamond$ CH sub-giants; $\square$
	CH giants; $\blacksquare$ C giant; $\blacktriangledown$
	halo CH giant;
	$\medbullet$ halo yellow symbiotic; $\triangle$ halo
	C-rich giant; $\blacktriangle$ halo C-rich sub-giant;
	$\triangledown$ halo N-rich dwarf,
	all previous gathered by \citet{2001ApJ...557..802B} ; 
	$\medcirc$ C stars from \citet{2002ApJ...579..817A};
	$\Diamondblack$ lead stars gathered by
	\citet{2006MmSAI..77..985B}.
	The observed data have an average error given by the size
	of the upper right cross in each plot.}
      \label{extr_stars}
   \end{figure*}

Extrinsic $s$-enhanced stars can be in any evolutionary
stage prior to the TP-AGB phase, but they are mainly giants
and main sequence stars.
Their absolute $s$-process abundances cannot be studied without
complete binary evolution models, but the abundance
ratios of $s$-process elements are not substantially affected
by binary processes. The [hs/ls] ratio
represents that of the former TP-AGB companion which
produced the $s$-process elements and it can be reasonably
approximated by taking into account the yields of single
stellar evolution models.
Fig. \ref{extr_stars} shows a comparison of our synthetic [hs/ls]
ratios to the extrinsic star data from \citet{2001ApJ...557..802B},
\citet{2002ApJ...580.1149A}, \citet{2002ApJ...579..817A} and
\citet{2003A&A...404..291V}.
Similar to our results for intrinsic stars, the spread observed
in these results is due to the range of initial
masses that we use, although,  in this case the mass range is wider
and includes AGB stars up to 7-8 M$_{\odot}$.
An AGB star of $M\gtrsim5$ M$_{\odot}$ has a very thin
intershell and a massive envelope, so
its surface [hs/ls] ratio varies smoothly between its value in the
intershell and the solar ratio. 

We focus first on the galactic disk (${\rm [Fe/H]}\gtrsim -1$)
stars. Fig. \ref{extr_stars} shows that
no single choice of $^{13}{\rm C}_{\rm eff}$ value 
can make our results fit all the observational data,
but almost all observations can be fitted with a range of values
$2/3\lesssim~^{13}{\rm C}_{\rm eff}\lesssim4/3$.
This is consistent with what we found for
the intrinsic $s$-enhanced stars (see $\S$\ref{intrinsic}).

For ${\rm [Fe/H]}\lesssim -1$, our synthetic [hs/ls] ratios are
relatively insensitive to changes in 
$^{13}{\rm C}_{\rm eff}$ because the hs-element synthesis
saturates, opening the way to the synthesis of lead.
There is some indication
that a smaller value of $^{13}{\rm C}_{\rm eff}$ is needed to
reproduce objects with [Fe/H] $\lesssim-2$ which have a high [hs/ls],
but even when using small $^{13}{\rm C}_{\rm eff}$
values [hs/ls] does not exceed about 1 dex. Thus we are unable
to explain the extreme low-metallicity objects with
${\rm [hs/ls]}\gtrsim1.2$.
Our synthetic calculations do not extend below [Fe/H]
$<-2.3$ because of the lack of detailed models with metallicity
below this value.

\subsubsection{Lead stars}
\begin{figure}
   \centering
   \includegraphics[width=0.5\textwidth]{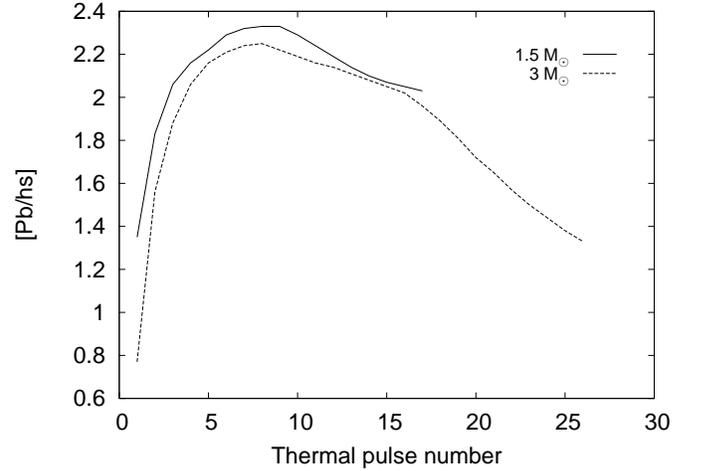}
      \caption{[Pb/hs] during the evolution of 1.5 M$_{\sun}$
    and 3 M$_{\sun}$ models of [Fe/H]=$-$2.3 with
    $^{13}$C$_{\rm eff}=1/3$.}
      \label{Fig:Pbhs}
   \end{figure}

\begin{figure*}
   \centering
   \includegraphics[width=\textwidth]{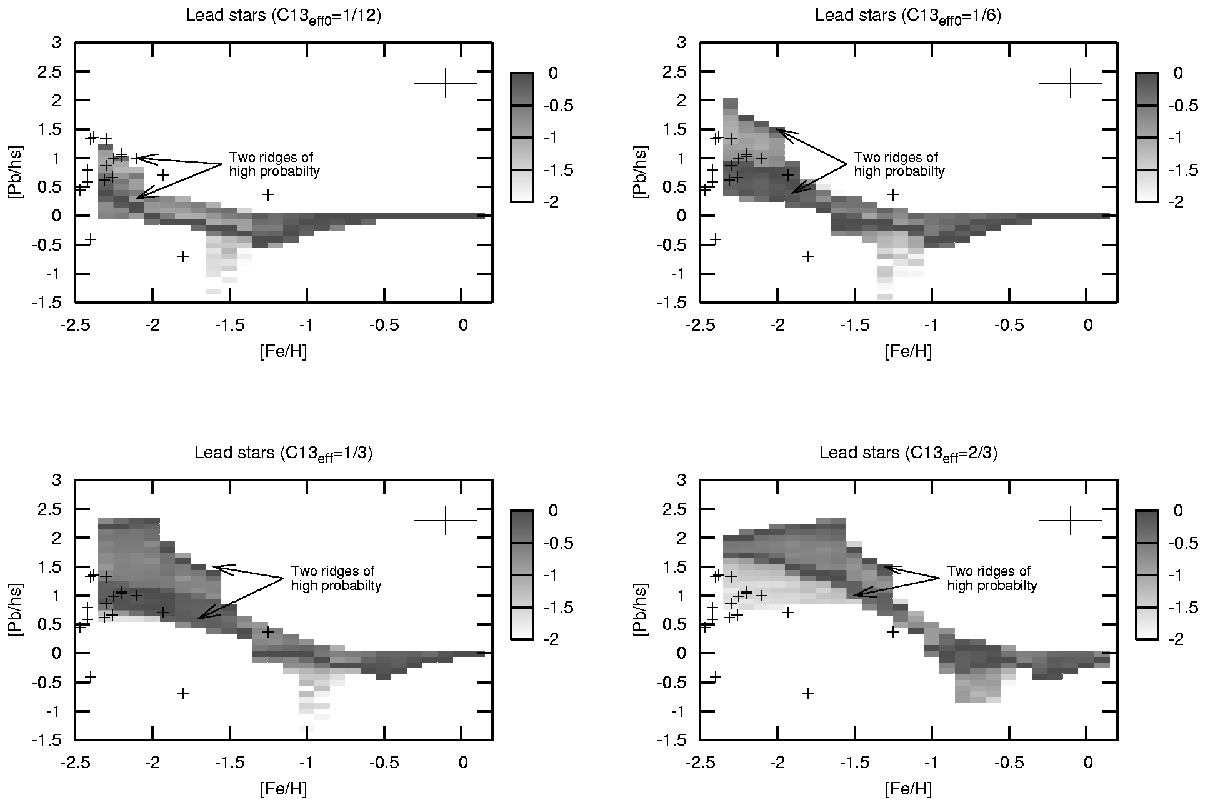}
      \caption{[Pb/hs] results from our population synthesis
	of lead stars, shown in the same
	way as in fig. \ref{extr_stars} and compared to the data
	gathered by \citet{2006MmSAI..77..985B}. The upper right
	cross shows the average error of the data points.}
      \label{lead_stars}
   \end{figure*}

Lead stars have an enhanced lead abundance and are found at
low metallicities. As discussed above, the [hs/ls] ratio
remains roughly constant at low metallicities
due to the saturation of the ls- and hs-element synthesis,
which indicates that lead is being synthesized
\citep{1998ApJ...497..388G}.
Consequently, the ratio of Pb to hs-elements
is sensitive to the choice of $^{13}{\rm C}_{\rm eff}$.
We use [Pb/hs] data gathered by \citet{2006MmSAI..77..985B}
to compare with our synthesis results.
In the [Pb/hs] vs. [Fe/H] plane (Fig. \ref{lead_stars})
we see that our results show a pattern that shifts in
metallicity if $^{13}{\rm C}_{\rm eff}$ is varied,
similar to the shift in [hs/ls] pattern at higher metallicities.
The reason for this behaviour is the change
in ratio of the number of free neutrons to the number of seed
nuclei, as explained in $\S$\ref{MS_S}.
We find that also for [Pb/hs] a spread naturally arises in our 
population synthesis results. This spread is mostly due to the fact
that, as discussed by \citet{1998ApJ...497..388G} (see their
Fig. 6), the neutron
exposure in the $^{13}$C pocket decreases with pulse number.
This effect is even more pronounced at lower metallicities, causing
the [Pb/hs] ratio to shift to much lower values as the evolution
proceeds. This is illustrated by the models computed by
\citet{2001ApJ...557..802B}, shown in Fig. \ref{Fig:Pbhs}.
As a consequence, there are two ridges of high
probability in the model results shown in Fig. \ref{lead_stars}.
One follows the higher [Pb/hs] values and corresponds to the
contribution of low-mass stars (about 1 M$_{\odot}$), which is high
due to their large IMF weight. The other ridge, which follows
intermediate values, is the contribution of stars with masses around
2.5 M$_{\odot}$. Despite their smaller IMF weight,
the ejecta of these stars are massive enough to make their
contribution comparable to that of the low-mass stars.
This effect is also visible in our
synthetic [hs/ls] results, e.g., in the top panels of
Fig. \ref{extr_stars}.
Most observations are fitted by the contribution of
$\sim2.5$ M$_{\odot}$ stars when choosing
$^{13}{\rm C}_{\rm eff}\approx1/3$ (lower left panel in Fig.
\ref{lead_stars}), except for two outliers which have
${\rm [Pb/hs]<0}$. These outliers, however, have large error
bars which extend to ${\rm [Pb/hs]\approx0}$
\citep{2003A&A...404..291V}. The upper left panel in Fig.
\ref{lead_stars} shows that a value of
$^{13}{\rm C}_{\rm eff}$ as small as $1/12$ 
still marginally fits the low-metallicity stars
(${\rm [Fe/H]}\lesssim -2$).
On the other hand, a choice of $^{13}{\rm C}_{\rm eff}\approx2/3$ 
results in too large [Pb/hs] values (lower right panel in Fig.
\ref{lead_stars}).
This result reaffirms our results for the extrinsic-star 
[hs/ls] values which suggest that
a somewhat smaller value of  $^{13}{\rm C}_{\rm eff}$ is needed
to fit low-metallicity halo stars compared to those in the
galactic disk. This is also consistent with what we found
in $\S$\ref{postagb}, regarding the object IRAS07134+1005.

As stated above our calculations do not extend below ${\rm [Fe/H]}<-2.3$,
but we notice that lead stars with lower metallicity could also
be explained if the trend shown by our models continues at lower
metallicities. This seems likely because [Pb/hs] does depend mostly
on $^{13}{\rm C}_{\rm eff}$ and not on other uncertain parameters
in our models, however, explaining those stars
at ${\rm [Fe/H]}<-2$ which show ${\rm [hs/ls]}\gtrsim1.2$ is still
a problem.

Finally, we note that \citet{2006MNRAS.368..305C} also presented a
possible solution to explain the spread of [Pb/hs] observed in
low-metallicity stars. These authors obtained a spread of neutron
exposures (and hence [Pb/hs]) in low-metallicity AGB stars due to
variations of the inter-pulse period in stars of different masses.
Their results are based on the assumption that the neutron
irradiation time is proportional to the inter-pulse period.
However, this assumption is incorrect since the neutron irradiation
time depends instead on the timescale at which the
${\rm ^{13}C}(\alpha,n)^{16}{\rm O}$ reactions occur.
At the temperature of $10^8$ K that \citet{2006MNRAS.368..305C}
use in their models, the timescale for $\alpha$ captures is of
the order of 300 yr (see Fig. 2 of \citealt{1998ApJ...497..388G}),
and this is independent of the inter-pulse period.
In our models, instead, the spread in [Pb/hs] derives naturally
from the fact that the neutron exposure changes with time,
especially at low metallicities.

\section{Conclusions and discussion}
\label{disc}

Based on the results of our population synthesis study, we find that
with the standard set of values for the free parameters in our AGB
models it is not possible to reproduce the observed heavy-element
abundances of $s$-enhanced stars.
The minimum core mass for TDU has to be reduced by at least 10 per
cent with respect to what theoretical models predict; low mass stars must
have a significant TDU efficiency and the $^{13}$C pocket mass must be
somewhat smaller than what is usually considered standard.  We also
find that the observations can be matched with a narrow range of
$^{13}{\rm C}_{\rm eff}$ values and that the mean $^{13}{\rm C}_{\rm
eff}$ value in this range apparently decreases with metallicity.

The observations in the metallicity range ${\rm [Fe/H]}\gtrsim-1$ are
well fit by $2/3\lesssim~^{13}{\rm C}_{\rm eff}\lesssim4/3$, i.e., a
spread of a factor of 2, while \citet{2001ApJ...557..802B} needed a
factor of $\sim20$. This is a consequence of the natural spread that
we find in the $s$-process element ratios caused mainly by the
truncation of AGB evolution due to the total loss of the stellar
envelope.  Stars with initial mass $M\lesssim1.5$ M$_{\odot}$ start
the TP-AGB phase with a relatively similar core-mass, but different
envelope mass. Consequently, a star with a certain low mass will
experience fewer thermal pulses than one with a slightly larger mass,
which affects their yields and composition during their life as a
TP-AGB star. A different choice of mass loss in our models, either
during the red giant branch or the AGB phase, would move the mass
range in which this effect takes place, but not eliminate it.

At ${\rm [Fe/H]}\lesssim-1$ a spread in [Pb/hs] arises because of
the shift of the neutron exposure.
Observations at these lower metallicities need a somewhat
smaller value $^{13}{\rm C}_{\rm eff}\approx1/3$, perhaps down to
$\approx1/12$.
This result may be due to the increasingly important
effect of $\alpha$-enhanced light neutron poisons at low
metallicities \citep{1999ARA&A..37..239B}, which needs to be
investigated in more detail.

It is reassuring that our conclusion of a small spread
in $^{13}{\rm C}_{\rm eff}$ is reached
independently by the study of isotopic $s$-process signatures in
pre-solar silicon carbide (SiC) grains from carbon stars. While
\citet{2003ApJ...586.1305L} needed a spread of $\sim24$ in the values of 
$^{13}{\rm C}_{\rm eff}$ to cover the SiC data, more recent
laboratory studies \citep{2006M&PS..XXB,2006M&PSA..41.5248M}, by
eliminating the effect of contamination of solar material, have also
reduced the spread of the $^{13}{\rm C}_{\rm eff}$ values needed to
cover the grain data to exactly the same range we have found here.
Although pre-solar grains can only give us information on the
$s$-process taking place in stars with ${\rm [Fe/H]}\gtrsim-1$,
such an independent check also gives us confidence in our
results at lower metallicities.

Now the question arises if the small spread that we find is
compatible with the possible mechanisms currently proposed for
the proton diffusion leading to the formation of the c13 pocket.
We note that 
semi-convection \citep{1988ApJ...333L..25H},
hydrodynamical overshooting \citep{1997A&A...324L..81H}
and gravity waves \citep{2003MNRAS.340..722D},
which are three out of the four proposed mechanisms,
produce a proton profile
with the number of protons varying continuously from the envelope
value to zero. The possible shapes of the proton profile where
discussed by \citet[see their Fig. 10]{2000A&A...362..599G}, who
concluded that ``the s-process predictions are only weakly
dependent on the shape of the H-profile''. Hence, from first
principle, we would not expect a wide range of neutron exposures
to occur in the current scenario. Future work should
quantitatively link the shape of the proton profile to the range
of $^{13}{\rm C}_{\rm eff}$ we have found in this study.
The fourth process that
has been proposed for the formation of the $^{13}{\rm C}$ neutron source
is rotation \citep{1999A&A...346L..37L, 2003ApJ...593.1056H}.
Also in this case the formation of the $^{13}{\rm C}$ pocket
starts from a continuous proton profile, however, current models
find that further mixing during the inter-pulse period completely
inhibits the s-process \citep{2003ApJ...593.1056H,2004A&A...415.1089S}.
Future work is need to ascertain this point.

To explain the existence of galactic post-AGB stars that are
$s$-process enhanced at [Fe/H] $\sim-1$ we find that the minimum
core mass for TDU must be 0.055-0.075 M$_{\odot}$ smaller than that
found in the models of \citet{2002PASA...19..515K} and
\citet{2004MNRAS.352..984S}. This is consistent with the findings of
several authors who studied the carbon-star luminosity functions of
the SMC and LMC (e.g., \citealt{1993A&A...267..410G},
\citealt{1999A&A...344..123M}, \citealt{2004MNRAS.350..407I} and
\citealt{2005MNRAS.356L...1S}) and sets a strong constraint on
future detailed evolutionary model results.

We find that to account for the fact that all 11 Zr-enhanced post-AGB
stars observed so far are also C enhanced, $\lambda_{\rm
min}\gtrsim0.2$ in stars of mass $M\lesssim1.5$ M$_{\odot}$. The
precise value of this lower limit may depend somewhat on our adopted
choices for mass loss, the AMR and the minimum core-mass for TDU. We
leave a detailed analysis of this dependence for future study.  
However, it is clear from our results that a value of $\lambda=0$ 
cannot reproduce the observations, and we are therefore
confident in concluding that stars of mass $M\lesssim1.5$ M$_{\odot}$
must experience a significant amount of TDU.

The amount of Zr enhancement in post-AGB stars is determined
by the amount of dredge up and the size of the $^{13}$C pocket.
Assuming $\lambda_{\rm min}=0.2$ indicates that in our models
the $^{13}$C pocket must have a mass of 1/40 of that of the
intershell to fit the observed data from Zr enhanced post-AGB
stars. This is somewhat smaller than the size of the $^{13}$C
pocket assumed in the models of \citet{1998ApJ...497..388G},
which is about 1/20 of the mass of the intershell.

Future detailed modelling of AGB stars will need to address the
constraints we have set here. We are currently extending our analysis
of extrinsic $s$-enhanced stars by means of \emph{binary} population
synthesis, i.e., by explicitly following the evolution and interaction
processes of populations of binary stars.

\begin{acknowledgements}
  The authors are deeply indebted to Roberto Gallino for providing
  the data from his detailed nucleosynthesis
  models and for insightful discussions and comments on this work.
  ABM thanks Maarten Reyniers and
  Hans van Winckel for useful communications, discussions and
  for providing data on post-AGB stars. ABM also thanks Selma de
  Mink for refreshing his knowledge of statistical analysis. RGI is
  supported by NWO. ML is supported by NWO (VENI fellow).
\end{acknowledgements}

\bibliography{bonacic_et_al_2007}

\begin{thebibliography}{52}
\expandafter\ifx\csname natexlab\endcsname\relax\def\natexlab#1{#1}\fi

\bibitem[{{Abia} {et~al.}(2001){Abia}, {Busso}, {Gallino}, {Dom{\'{\i}}nguez},
  {Straniero}, \& {Isern}}]{2001ApJ...559.1117A}
{Abia}, C., {Busso}, M., {Gallino}, R., {et~al.} 2001, \apj, 559, 1117

\bibitem[{{Abia} {et~al.}(2002){Abia}, {Dom{\'{\i}}nguez}, {Gallino}, {Busso},
  {Masera}, {Straniero}, {de Laverny}, {Plez}, \&
  {Isern}}]{2002ApJ...579..817A}
{Abia}, C., {Dom{\'{\i}}nguez}, I., {Gallino}, R., {et~al.} 2002, \apj, 579,
  817

\bibitem[{{Anders} \& {Grevesse}(1989)}]{1989GeCoA..53..197A}
{Anders}, E. \& {Grevesse}, N. 1989, \gca, 53, 197

\bibitem[{{Aoki} {et~al.}(2002){Aoki}, {Ryan}, {Norris}, {Beers}, {Ando}, \&
  {Tsangarides}}]{2002ApJ...580.1149A}
{Aoki}, W., {Ryan}, S.~G., {Norris}, J.~E., {et~al.} 2002, \apj, 580, 1149

\bibitem[{{Barzyk} {et~al.}(2006){Barzyk}, {Savina}, {Davis}, {Gallino},
  {Gyngard}, {Amari}, {Zinner}, {Pellin}, {Lewis}, \&
  {Clayton}}]{2006M&PS..XXB}
{Barzyk}, J.~G., {Savina}, M.~R., {Davis}, A.~M., {et~al.} 2006, submitted to
  Meteoritics \& Planetary Science

\bibitem[{{Bisterzo} {et~al.}(2006){Bisterzo}, {Gallino}, {Straniero}, {Ivans},
  {K{\"a}ppeler}, \& {Aoki}}]{2006MmSAI..77..985B}
{Bisterzo}, S., {Gallino}, R., {Straniero}, O., {et~al.} 2006, Memorie della
  Societa Astronomica Italiana, 77, 985

\bibitem[{{Boothroyd} \& {Sackmann}(1988)}]{1988ApJ...328..632B}
{Boothroyd}, A.~I. \& {Sackmann}, I.-J. 1988, \apj, 328, 632

\bibitem[{{Busso} {et~al.}(2001){Busso}, {Gallino}, {Lambert}, {Travaglio}, \&
  {Smith}}]{2001ApJ...557..802B}
{Busso}, M., {Gallino}, R., {Lambert}, D.~L., {Travaglio}, C., \& {Smith},
  V.~V. 2001, \apj, 557, 802

\bibitem[{{Busso} {et~al.}(1999){Busso}, {Gallino}, \&
  {Wasserburg}}]{1999ARA&A..37..239B}
{Busso}, M., {Gallino}, R., \& {Wasserburg}, G.~J. 1999, \araa, 37, 239

\bibitem[{{Carraro} {et~al.}(1996){Carraro}, {Girardi}, {Bressan}, \&
  {Chiosi}}]{1996A&A...305..849C}
{Carraro}, G., {Girardi}, L., {Bressan}, A., \& {Chiosi}, C. 1996, \aap, 305,
  849

\bibitem[{{Cui} \& {Zhang}(2006)}]{2006MNRAS.368..305C}
{Cui}, W. \& {Zhang}, B. 2006, \mnras, 368, 305

\bibitem[{{Denissenkov} \& {Tout}(2003)}]{2003MNRAS.340..722D}
{Denissenkov}, P.~A. \& {Tout}, C.~A. 2003, \mnras, 340, 722

\bibitem[{{Frost} \& {Lattanzio}(1996)}]{1996ApJ...473..383F}
{Frost}, C.~A. \& {Lattanzio}, J.~C. 1996, \apj, 473, 383

\bibitem[{{Gallino} {et~al.}(1998){Gallino}, {Arlandini}, {Busso}, {Lugaro},
  {Travaglio}, {Straniero}, {Chieffi}, \& {Limongi}}]{1998ApJ...497..388G}
{Gallino}, R., {Arlandini}, C., {Busso}, M., {et~al.} 1998, \apj, 497, 388

\bibitem[{{Giridhar} \& {Arellano Ferro}(2005)}]{2005A&A...443..297G}
{Giridhar}, S. \& {Arellano Ferro}, A. 2005, \aap, 443, 297

\bibitem[{{Goriely} \& {Mowlavi}(2000)}]{2000A&A...362..599G}
{Goriely}, S. \& {Mowlavi}, N. 2000, \aap, 362, 599

\bibitem[{{Groenewegen} \& {de Jong}(1993)}]{1993A&A...267..410G}
{Groenewegen}, M.~A.~T. \& {de Jong}, T. 1993, \aap, 267, 410

\bibitem[{{Herwig}(2005)}]{2005ARA&A..43..435H}
{Herwig}, F. 2005, \araa, 43, 435

\bibitem[{{Herwig} {et~al.}(1997){Herwig}, {Bloecker}, {Schoenberner}, \& {El
  Eid}}]{1997A&A...324L..81H}
{Herwig}, F., {Bloecker}, T., {Schoenberner}, D., \& {El Eid}, M. 1997, \aap,
  324, L81

\bibitem[{{Herwig} {et~al.}(2003){Herwig}, {Langer}, \&
  {Lugaro}}]{2003ApJ...593.1056H}
{Herwig}, F., {Langer}, N., \& {Lugaro}, M. 2003, \apj, 593, 1056

\bibitem[{{Hollowell} \& {Iben}(1988)}]{1988ApJ...333L..25H}
{Hollowell}, D. \& {Iben}, I.~J. 1988, \apjl, 333, L25

\bibitem[{{Hurley} {et~al.}(2000){Hurley}, {Pols}, \&
  {Tout}}]{2000MNRAS.315..543H}
{Hurley}, J.~R., {Pols}, O.~R., \& {Tout}, C.~A. 2000, \mnras, 315, 543

\bibitem[{{Iben}(1977)}]{1977ApJ...217..788I}
{Iben}, Jr., I. 1977, \apj, 217, 788

\bibitem[{{Iben} \& {Renzini}(1983)}]{1983ARA&A..21..271I}
{Iben}, Jr., I. \& {Renzini}, A. 1983, \araa, 21, 271

\bibitem[{{Izzard} {et~al.}(2006){Izzard}, {Dray}, {Karakas}, {Lugaro}, \&
  {Tout}}]{2006A&A...460..565I}
{Izzard}, R.~G., {Dray}, L.~M., {Karakas}, A.~I., {Lugaro}, M., \& {Tout},
  C.~A. 2006, \aap, 460, 565

\bibitem[{{Izzard} {et~al.}(2004){Izzard}, {Tout}, {Karakas}, \&
  {Pols}}]{2004MNRAS.350..407I}
{Izzard}, R.~G., {Tout}, C.~A., {Karakas}, A.~I., \& {Pols}, O.~R. 2004,
  \mnras, 350, 407

\bibitem[{{Jorissen} {et~al.}(1993){Jorissen}, {Frayer}, {Johnson}, {Mayor}, \&
  {Smith}}]{1993A&A...271..463J}
{Jorissen}, A., {Frayer}, D.~T., {Johnson}, H.~R., {Mayor}, M., \& {Smith},
  V.~V. 1993, \aap, 271, 463

\bibitem[{{Karakas} {et~al.}(2002){Karakas}, {Lattanzio}, \&
  {Pols}}]{2002PASA...19..515K}
{Karakas}, A.~I., {Lattanzio}, J.~C., \& {Pols}, O.~R. 2002, \pasa, 19, 515

\bibitem[{{Kroupa} {et~al.}(1993){Kroupa}, {Tout}, \&
  {Gilmore}}]{1993MNRAS.262..545K}
{Kroupa}, P., {Tout}, C.~A., \& {Gilmore}, G. 1993, \mnras, 262, 545

\bibitem[{{Langer} {et~al.}(1999){Langer}, {Heger}, {Wellstein}, \&
  {Herwig}}]{1999A&A...346L..37L}
{Langer}, N., {Heger}, A., {Wellstein}, S., \& {Herwig}, F. 1999, \aap, 346,
  L37

\bibitem[{{Lattanzio}(1989)}]{1989ApJ...344L..25L}
{Lattanzio}, J.~C. 1989, \apjl, 344, L25

\bibitem[{{Lugaro} {et~al.}(2003){Lugaro}, {Herwig}, {Lattanzio}, {Gallino}, \&
  {Straniero}}]{2003ApJ...586.1305L}
{Lugaro}, M., {Herwig}, F., {Lattanzio}, J.~C., {Gallino}, R., \& {Straniero},
  O. 2003, \apj, 586, 1305

\bibitem[{{Marhas} {et~al.}(2006){Marhas}, {Hoppe}, \&
  {Ott}}]{2006M&PSA..41.5248M}
{Marhas}, K.~K., {Hoppe}, P., \& {Ott}, U. 2006, Meteoritics \& Planetary
  Science, Vol.~41, Supplement, Proceedings of 69th Annual Meeting of the
  Meteoritical Society, held August 6-11, 2006 in Zurich, Switzerland., p.5248,
  41, 5248

\bibitem[{{Marigo} {et~al.}(1999){Marigo}, {Girardi}, \&
  {Bressan}}]{1999A&A...344..123M}
{Marigo}, P., {Girardi}, L., \& {Bressan}, A. 1999, \aap, 344, 123

\bibitem[{{Merrill}(1952)}]{1952ApJ...116...21M}
{Merrill}, S.~P.~W. 1952, \apj, 116, 21

\bibitem[{{Mowlavi}(1999)}]{1999A&A...344..617M}
{Mowlavi}, N. 1999, \aap, 344, 617

\bibitem[{{Paczy{\'n}ski}(1970)}]{1970AcA....20...47P}
{Paczy{\'n}ski}, B. 1970, \aca, 20, 47

\bibitem[{{Pols} {et~al.}(1998){Pols}, {Schroder}, {Hurley}, {Tout}, \&
  {Eggleton}}]{1998MNRAS.298..525P}
{Pols}, O.~R., {Schroder}, K.-P., {Hurley}, J.~R., {Tout}, C.~A., \&
  {Eggleton}, P.~P. 1998, \mnras, 298, 525

\bibitem[{{Pont} \& {Eyer}(2004)}]{2004MNRAS.351..487P}
{Pont}, F. \& {Eyer}, L. 2004, \mnras, 351, 487

\bibitem[{{Reimers}(1975)}]{1975psae.book..229R}
{Reimers}, D. 1975, {Circumstellar envelopes and mass loss of red giant stars}
  (Problems in stellar atmospheres and envelopes.), 229--256

\bibitem[{{Reyniers} {et~al.}(2004){Reyniers}, {Van Winckel}, {Gallino}, \&
  {Straniero}}]{2004A&A...417..269R}
{Reyniers}, M., {Van Winckel}, H., {Gallino}, R., \& {Straniero}, O. 2004,
  \aap, 417, 269

\bibitem[{{Siess} {et~al.}(2004){Siess}, {Goriely}, \&
  {Langer}}]{2004A&A...415.1089S}
{Siess}, L., {Goriely}, S., \& {Langer}, N. 2004, \aap, 415, 1089

\bibitem[{{Smith} \& {Lambert}(1990)}]{1990ApJS...72..387S}
{Smith}, V.~V. \& {Lambert}, D.~L. 1990, \apjs, 72, 387

\bibitem[{{Stancliffe} {et~al.}(2005){Stancliffe}, {Izzard}, \&
  {Tout}}]{2005MNRAS.356L...1S}
{Stancliffe}, R.~J., {Izzard}, R.~G., \& {Tout}, C.~A. 2005, \mnras, 356, L1

\bibitem[{{Stancliffe} {et~al.}(2004){Stancliffe}, {Tout}, \&
  {Pols}}]{2004MNRAS.352..984S}
{Stancliffe}, R.~J., {Tout}, C.~A., \& {Pols}, O.~R. 2004, \mnras, 352, 984

\bibitem[{{Straniero} {et~al.}(1997){Straniero}, {Chieffi}, {Limongi}, {Busso},
  {Gallino}, \& {Arlandini}}]{1997ApJ...478..332S}
{Straniero}, O., {Chieffi}, A., {Limongi}, M., {et~al.} 1997, \apj, 478, 332

\bibitem[{{Straniero} {et~al.}(2003){Straniero}, {Dom{\'{\i}}nguez},
  {Cristallo}, \& {Gallino}}]{2003PASA...20..389S}
{Straniero}, O., {Dom{\'{\i}}nguez}, I., {Cristallo}, R., \& {Gallino}, R.
  2003, \pasa, 20, 389

\bibitem[{{Travaglio} {et~al.}(1999){Travaglio}, {Galli}, {Gallino}, {Busso},
  {Ferrini}, \& {Straniero}}]{1999ApJ...521..691T}
{Travaglio}, C., {Galli}, D., {Gallino}, R., {et~al.} 1999, \apj, 521, 691

\bibitem[{{van Eck} {et~al.}(2003){van Eck}, {Goriely}, {Jorissen}, \&
  {Plez}}]{2003A&A...404..291V}
{van Eck}, S., {Goriely}, S., {Jorissen}, A., \& {Plez}, B. 2003, \aap, 404,
  291

\bibitem[{{van Winckel}(2003)}]{2003ARA&A..41..391V}
{van Winckel}, H. 2003, \araa, 41, 391

\bibitem[{{Van Winckel} \& {Reyniers}(2000)}]{2000A&A...354..135V}
{Van Winckel}, H. \& {Reyniers}, M. 2000, \aap, 354, 135

\bibitem[{{Vassiliadis} \& {Wood}(1993)}]{1993ApJ...413..641V}
{Vassiliadis}, E. \& {Wood}, P.~R. 1993, \apj, 413, 641

\end{thebibliography}

\begin{appendix}
\section{Details of the synthetic TP-AGB model}
\label{apa}

\citet[][hereafter H00]{2000MNRAS.315..543H} developed a comprehensive
synthetic evolution code for single stellar evolution,
based on detailed models of \citet[][hereafter P98]{1998MNRAS.298..525P}.
The P98 models include convective core overshooting, but do not
undergo thermal pulses during the AGB phase.
To model TP-AGB stars, H00 apply the same luminosity-core mass
relation as in the early (E-) AGB phase.
The TP-AGB section of the H00 code was improved by
\citet[][hereafter I04]{2004MNRAS.350..407I} with a synthetic code
based on the detailed model calculations by
\citet[][hereafter K02]{2002PASA...19..515K} which undergo thermal
pulses. The latest version \citep{2006A&A...460..565I}
also follows the nucleosynthesis of many $s$-process
isotopes based on the models of \citet{1998ApJ...497..388G}.

However, unlike the P98 detailed models, the K02 detailed
models do not include convective overshooting. This means that the
transition from the E-AGB to the TP-AGB phase is not self-consistent in
the I04 code, leading to discontinuities in the evolution of the stellar
core mass, the luminosity and the radius of the AGB star.
To overcome these problems we have made modifications to the I04
code, which are explained in detail below.

\subsection{TP-AGB initial H-depleted core mass}
   \begin{figure}
   \centering
   \includegraphics[width=0.5\textwidth]{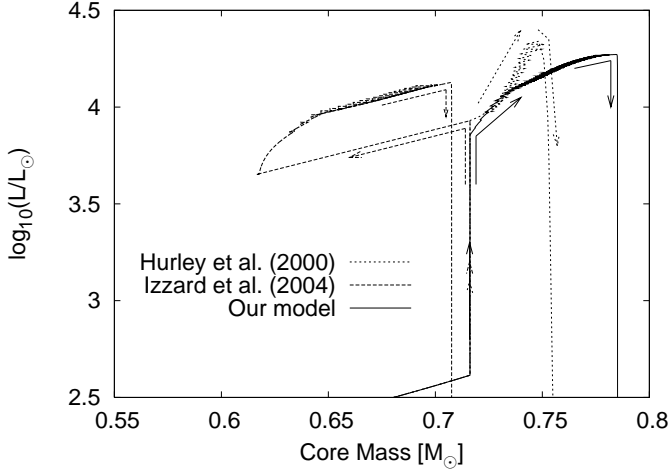}
      \caption{Luminosity$-$core-mass relation of a 3 M$_{\odot}$
	star with metallicity $Z=0.01$. The arrows show the way in which
	the core evolves in time. During the E-AGB phase, the
	hydrogen-exhausted core mass remains constant at 0.716
	M$_{\odot}$. In our model the
	core mass at the start of the TP-AGB is the same as in the H02 model,
	which is greater than in the I04 model.}
      \label{model_comp_lmc}
   \end{figure}

For an AGB star to enter the TP-AGB phase, its helium intershell
mass must be small enough to undergo
a flash. This occurs at the end of the E-AGB
phase when the helium burning shell almost reaches the hydrogen
burning shell.
In our code the H-depleted core mass at the beginning
of the TP-AGB
phase, $M_{\rm c,1TP}$, is given by the mass of the H-depleted core
at the base of the AGB, $M_{\rm c,BAGB}$. However, stars with
$M_{\rm c,BAGB} \sim0.8$M$_{\odot}$ and larger undergo second
dredge-up by the end of the E-AGB phase, which makes
$M_{\rm c,1TP}<M_{\rm c,BAGB}$.

When convective overshooting is taken into account, the value of
$M_{\rm c,1TP}$ for a star with a given initial mass is larger than that
of a star with the same initial mass without overshooting.
The effects of overshooting become
important in AGB stars of $M\gtrsim 2.5$M$_{\odot}$\footnote{The effects
of overshooting are also strong in lower-mass stars on the main sequence
and the horizontal branch phase, but are largely wiped out during the red
giant branch phase \citep{1998MNRAS.298..525P}.}, having a direct
impact on the subsequent evolution.

We have improved the linear relation between $M_{\rm c,BAGB}$ and
$M_{\rm c,1TP}$ from H00 (denoted as $M_{\rm c,DU}$ in that paper) with a
quadratic one, based on data from the P98 models (hereafter OV):
\begin{equation}
\begin{split}
M_{\rm c,1TP,OV}=~&9.292\times10^{-2}M_{\rm c,BAGB}
\times (M_{\rm c,BAGB}+1.983)\\&+5.865\times10^{-1},
\end{split}
\end{equation}
in solar units.
This fit is
not very good for stars with $M\lesssim1.5$M$_{\odot}$,
but in these the effects of overshooting
become negligible and the fit to the K02 models ($M_{\rm c,1TP,K02}$)
is appropriate, so we use instead
\begin{equation}
\begin{split}
M_{\rm c,1TP}=~&\rm{max}\Big[\rm{~min}\big(M_{\rm c,1TP,OV},M_{\rm c,BAGB}\big),\\
&M_{\rm c,1TP,K02}\Big].
\end{split}
\end{equation}
Fig. \ref{model_comp_lmc} shows a luminosity$-$core-mass plot
where the core-mass prescriptions of the different synthetic models
can be compared.

\subsection{CO core mass and luminosity on E-AGB phase}

   \begin{figure}
   \centering
   \includegraphics[width=0.5\textwidth]{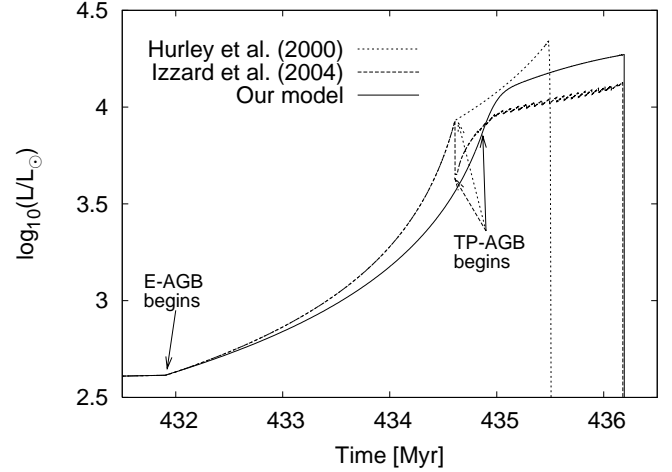}
   \caption{Evolution of luminosity in time of the same star as in
     fig. \ref{model_comp_lmc}. The E-AGB evolution is slightly
     modified in our model and the discontinuity in luminosity
     is fixed.}
   \label{model_comp_lt}
   \end{figure}

As in H00, we model the luminosity during the E-AGB phase with 
a power-law relation of the form $L_{\rm E-AGB}=l_1M_{\rm CO}^{l_2}$, where
$M_{\rm CO}$ is the mass of the CO core during the E-AGB phase.
The coefficients $l_1$ and $l_2$ are found from
the values of $L_{\rm E-AGB}$ and $M_{\rm CO}$ at the beginning and at the
end of the E-AGB phase.

For the beginning we use Eq. (56) from H00 for the luminosity
at the base of the AGB and the following relation
for $M_{\rm CO}$ at the base of the AGB
\begin{equation}
M_{\rm CO,BAGB}=(a_1M^{a_2}+a_3)^{0.25},
\end{equation} 
which fits the data from the P98 detailed models,
expressed in solar units. The coefficients depend on metallicity
and are calculated with the formula
\begin{equation}
a_i=a_{i,1}+{\rm [Fe/H]}{\big(}a_{i,2}+{\rm [Fe/H]}(a_{i,3}+{\rm [Fe/H]}a_{i,4}){\big)}, 
\label{mco_ai_coefs}
\end{equation}
where $a_{i,j}$ are the dimensionless coefficients listed in table
\ref{mco_aij_table}.

\begin{table}
\caption{Coefficients $a_{i,j}$ for Eq. (\ref{mco_ai_coefs}).}
\label{mco_aij_table}
\begin{center}
\begin{tabular}{c|cccc}
$a_{i,j}$ & $j=1$ & $j=2$ & $j=3$ & $j=4$\\
\hline
$i=1$ & 1.287$\times10^{-5}$ & -3.709$\times10^{-5}$ & 6.535$\times10^{-5}$ & 1.607$\times10^{-5}$\\
$i=2$ & 5.881 & 8.617$\times10^{-1}$ & 3.410$\times10^{-1}$ & 6.039$\times10^{-2}$\\
$i=3$ & 4.878$\times10^{-3}$ & -1.764$\times10^{-4}$ & 1.986$\times10^{-3}$ & 6.635$\times10^{-4}$\\
\end{tabular}
\end{center}
\end{table}

For a smooth transition to the TP-AGB phase, we require
\begin{equation}
M_{\rm CO,end}=M_{\rm c,1TP}~~~{\rm and}~~~L_{\rm E-AGB,end}=L_{\rm 1TP},
\end{equation}
where $M_{\rm CO,end}$ and $L_{\rm E-AGB,end}$ are
the CO core mass and the luminosity at the end of the E-AGB phase,
respectively. $L_{\rm 1TP}$ is the luminosity at the beginning of the
TP-AGB phase and is discussed in $\S$\ref{Ltp1}. A comparison
between the time evolution of the luminosity
described by the different synthetic models is shown in fig.
\ref{model_comp_lt}.

   \begin{figure*}
   \centering
   \includegraphics[width=\textwidth]{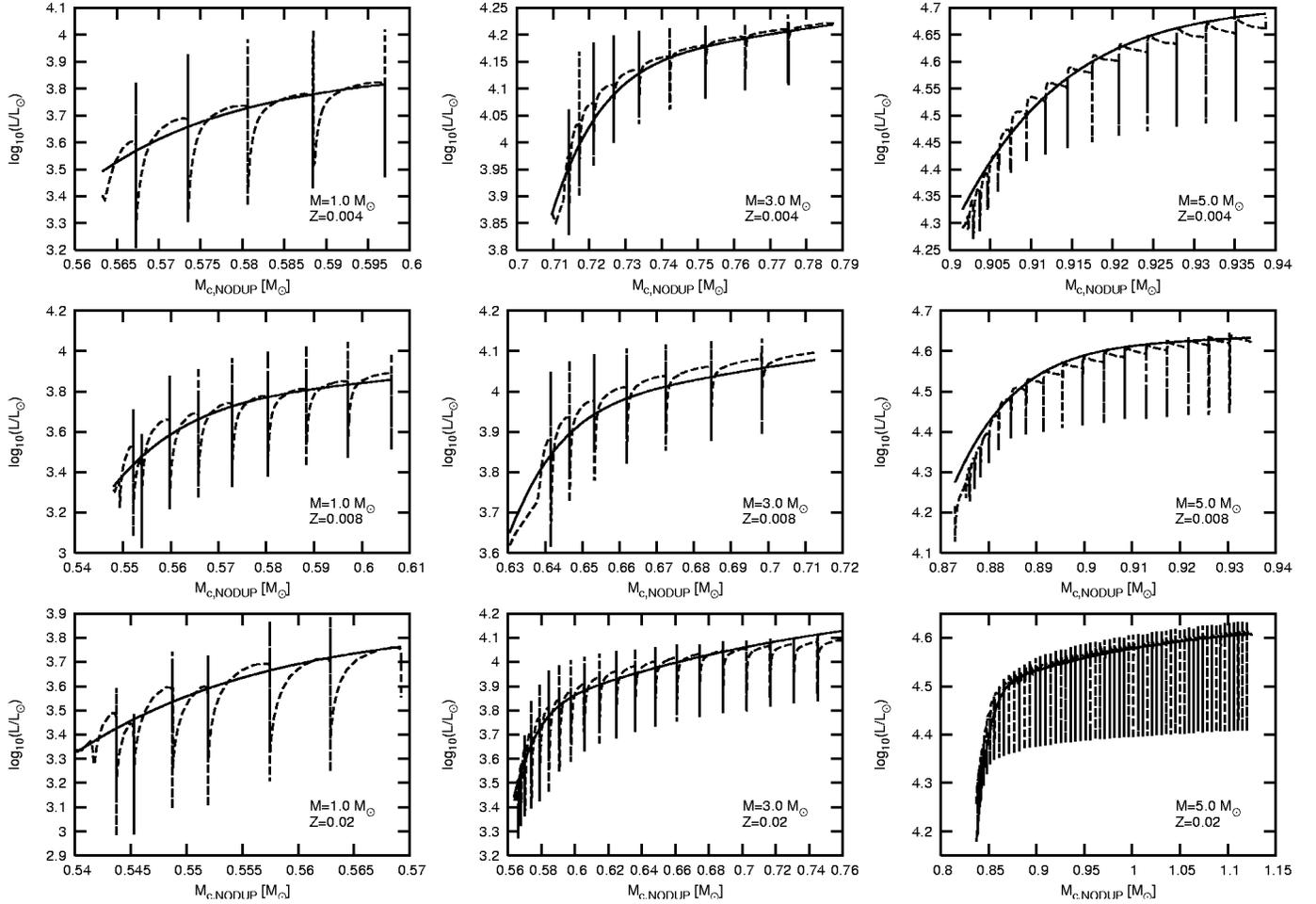}
   \caption{Examples of our analytic fit (solid lines) compared to
     the data from the detailed models (dashed lines) from
     \cite{2004MNRAS.352..984S} for different stellar masses and
     metallicities, as indicated in the panels.}
   \label{lums_comp}
   \end{figure*}

\subsection{Luminosity during TP-AGB phase}
\label{Ltp1}

Following \citet{1970AcA....20...47P}, it has become usual
practice to apply a linear luminosity$-$core-mass relation in
synthetic AGB models.
However, this relation fails in the presence of dredge-up and hot
bottom burning. This becomes a problem when comparing models
which include overshooting to models that do not: two TP-AGB stars with
the same core mass, but different envelope mass, do not behave in
the same way. To overcome this we design
a luminosity relation which depends on both the
core mass and the envelope mass of the star. It
successfully reproduces both the P98 models which include
overshooting and models calculated with the same code used by P98
which do not include overshooting.
\citet[][hereafter S04]{2004MNRAS.352..984S} employed a modified
version of the code used by P98 and computed detailed models which do
not include overshooting, but do undergo thermal pulses.
In an attempt to make our rapid TP-AGB
models more self-consistent we have modified our luminosity
relation to fit the S04 models, assuming that with the inclusion
of overshooting the S04 models are modified in the same way
as those of P98. In addition to this, we have found that in
the presence of dredge-up the results of
the detailed stellar evolution models are better represented by
relating the luminosity to $M_{\rm c,NODUP}$, the core mass
as it would be in the absence of dredge-up:
\begin{equation}
\begin{split}
\log_{10}(L_{\rm c,NODUP})=~&\Bigg(\log_{10}(15300+9440Z)\\
&+\frac{3.45~\log_{10}
(M_{\rm c,NODUP}+0.1)}
{3M_{\rm c,NODUP}-M_{\rm c,1TP}}\Bigg)\\
&\times\left(\frac{M_{\rm c,1TP}}{M_{\rm c,NODUP}}\right)^{0.045}+0.08,
\end{split}
\label{lc}
\end{equation}
in solar units and where $Z$ is the metallicity. 
In stars with a massive convective envelope
($M_{\rm env}\gtrsim2.5$ M$_{\odot}$),
the temperature at the bottom of the envelope is high enough for
hot bottom burning (HBB), which enhances the luminosity of the star
by a factor $f_{\rm HBB}$ due to the enhanced hydrogen abundance in
the H-burning shell.
The expression
\begin{equation}
\label{le0}
\begin{split}
\log_{10}(f_{\rm HBB})=~&\max\Bigg\{1.3~\log_{10}\left[
\frac{M_{\rm env,1TP}
(1-0.53~{\rm [Fe/H]})}{M_{\rm env,HBB}}\right]\\&-0.08~,~0\Bigg\}
\end{split}
\end{equation}
fits $f_{\rm HBB}\times L_{\rm c,NODUP}$ to the luminosity of the
S04 models, where $M_{\rm env,1TP}$ is
the mass of the envelope at the beginning of the TP-AGB and we have
set the threshold mass for the onset of HBB
$M_{\rm env,HBB}=2.4$M$_{\odot}$.

The S04 models have no mass loss, so we have to use a different set
of models to fit the
behaviour of the luminosity in terms of the loss of the envelope mass.
Based on the K02 models the contribution of the envelope
to the luminosity can be modelled as
\begin{equation}
\label{le}
\begin{split}
\log_{10}(f_{\rm env})=~&\max\Bigg\{\log_{10}(f_{\rm HBB})+\\
&\left.0.46~\min\left[\log_{10}\left(\frac{M_{\rm env}}
{M_{\rm env,1TP}}\right),0\right]~,~0\right\}
\end{split}
\end{equation}

The expression $f_{\rm env}\times L_{\rm c,NODUP}$
matches the asymptotic behaviour of the luminosity
observed in the S04 detailed models, which is reached after several
pulses. To model the first few pulses
we write the luminosity as
\begin{equation}
\label{l}
L=\left[\frac{L_{\rm c,NODUP}\times f_{\rm env}}
{L_{\rm TP1}}\right]^{f}L_{\rm TP1},
\end{equation}
where $L_{\rm TP1}$ is fitted to the S04 models with the expression
\begin{equation}
\label{ltp1}
\log_{10}(L_{\rm TP1})=-1.8427M_{\rm c,TP1}^2+5.3861M_{\rm c,TP1}+0.97823
\end{equation}
and
\begin{equation}
\label{ft}
f=1- \left[1-\frac{M_{\rm c,NODUP}-M_{\rm c,1TP}}{0.075}\right]^{25/3}
\end{equation}
accounts for the luminosity growth during the early
pulses to its asymptotic value $L_{\rm c,NODUP}\times f_{\rm env}$.
In Eqs. (\ref{ltp1}) and (\ref{ft}) all quantities are expressed in solar
units. Fig. \ref{lums_comp} shows a comparison between our fit and the
S04 detailed models. In the worst cases, our fits have an error smaller
than 10\% in luminosity, which is less than the difference between the
S04 and K02 detailed models.

\subsection{Radius during TP-AGB phase}

We use the H00 radius-luminosity relation, given that
it accurately fits the S04 models, but we have slightly modified
the interpolation procedure.
The H00 radius-luminosity relation (Eq. (74) in that paper) was
designed to fit the P98
models at seven different metallicities in the following way:
\begin{equation}
R_{\rm AGB}=A(L^{b_1}+b_2L^{b_3}),
\label{rad_lum_eq}
\end{equation}  
where $L$ is the luminosity, $A$ depends on the mass, and $b_i$, on the
metallicity. For intermediate metallicities the radius was calculated
after interpolating linearly in $Z$ between the coefficients that
fit these seven metallicities. Some of these coefficients are exponents
and we found the results for interpolated metallicities to behave
erratically. We therefore interpolate linearly in $Z$
between the radius values calculated with Eq. \ref{rad_lum_eq}
for the adjacent known metallicities,
which is in much better agreement with detailed models at intermediate
metallicities calculated with the same code as used by P98.

\end{appendix}

\end{document}